\documentclass[%
 reprint,
 onecolumn,
 amsmath,amssymb,
 aps,
]{revtex4-1}

\usepackage{graphicx}% Include figure files
\usepackage{dcolumn}% Align table columns on decimal point
\usepackage{bm}% bold math
\usepackage{color}
\usepackage{colordvi}  
\definecolor{Red}{rgb}{1,0,0}

\newcommand{\bra}[1]{\left\langle #1 \right|}
\newcommand{\ket}[1]{\left| #1 \right\rangle}

\begin{document}

\preprint{APS/123-QED}

\title{Generation of non-classical light using semiconductor quantum dots}% Force line breaks with \\

\author{Rahul Trivedi}
 \affiliation{E. L. Ginzton Laboratory, Stanford University, Stanford, California 94305, USA}
\author{Kevin A. Fischer}
 \affiliation{E. L. Ginzton Laboratory, Stanford University, Stanford, California 94305, USA}
\author{Jelena Vu\v{c}kovi\'{c}}%
 \affiliation{E. L. Ginzton Laboratory, Stanford University, Stanford, California 94305, USA}%

\author{Kai M\"uller}
 \email{kai.mueller@wsi.tum.de}
 \affiliation{Walter Schottky Institut and Physik Department, Technische Universit\"at M\"unchen, 85748 Garching, Germany}%

\date{\today}% It is always \today, today,
             %  but any date may be explicitly specified

\begin{abstract}
Sources of non-classical light are of paramount importance for future applications in quantum science and technology such as quantum communication, quantum computation and simulation, quantum sensing and quantum metrology. In this review we discuss the fundamentals and recent progress in the generation of single photons, entangled photon pairs and photonic cluster states using semiconductor quantum dots. Specific fundamentals which are discussed are a detailed quantum description of light, properties of semiconductor quantum dots and light-matter interactions. This includes a framework for the dynamic modelling of non-classical light generation and two-photon interference. Recent progress will be discussed in the generation of non-classical light for off-chip applications as well as implementations for scalable on-chip integration.
\end{abstract}

\maketitle

%\tableofcontents

\section{Introduction}
Non-classical states of light such as single photons and entangled photons are key ingredients for photonic quantum technologies with applications in quantum communication, quantum computation, quantum simulation, quantum sensing and quantum metrology \cite{Gisin2002, Gisin2007, OBrien2007, Giovannetti2004}. Over the last two decades, semiconductor quantum dots (QDs) have established themselves as a very promising material platform for sources of such non-classical light \cite{michler}. This results from their strong interband transitions with near-unity quantum efficiency, almost exclusive emission into the zero phonon line (ZPL) and nearly transform-limited linewidth \cite{Matthiesen2012, Prechtel2013, Hansom2014, Kuhlmann2015}. Moreover, the used III-V semiconductor materials allow to embed the QDs into opto-electronic devices which allows to electrically control the properties of the QDs, such as the emission wavelength \cite{Warburton2000}. Fabricating nanophotonic structures such as resonators and waveguides around QDs allows to further enhance the light-matter interaction in order to speed-up operational rates and to efficiently collect the generated non-classical light off-chip or route in on-chip for integrated quantum photonic circuits \cite{Pelton2002}. For example, a solitary high-quality QD source was recently used in exciting demonstrations to create a train of single photons, which were temporally multiplexed to the input of a Boson Sampler \cite{Loredo2017, Wang2017}.

Several books have already been written about semiconductor QDs \cite{michler} and multiple nice review have been recently been published, for example on the generation of single photons \cite{Aharonovich2016, Senellart2017}, entangled photon pairs \cite{Orieux2017, Huber2018review} or interfacing QDs with photonic nanostructures \cite{Lodahl2015, Lodahl2018}. Nevertheless, we believe that the community will  benefit from this review due to its combination of a thorough quantum optical theoretical background with a summary of the progress in all the fields dealing with QD-based sources of non-classical light such as single photons, entangled photon pairs and cluster states. In section \ref{section:fundamentals} we will discuss all relevant fundamentals, including a thorough quantum optical description of light and the light-matter interaction. This also includes a framework that can be used to simulate the quantum optical properties of non-classical light generated by many different schemes. Then, in sections \ref{section:single_photon}-\ref{section:graph} we review the recent progress in the generation of single photons, entangled photon pairs and cluster states using semiconductor QDs.

\section{Fundamentals}
\label{section:fundamentals}

\subsection{Quantum description of light}
A fully quantum description of electromagnetic fields is important for accurately describing the properties of light emitted by systems such as quantum dots. Quantization of an electromagnetic field can be performed by applying the second quantization principle to the classical Maxwell's equations. In the quantum description of electromagnetic fields, the electric field at each point $\textbf{x}$ in space is treated as an observable whose statistics are captured by a hermitian operator $\hat{\textbf{E}}(\textbf{x})$. The dynamics of the electromagnetic fields are described by a hamiltonian $\hat{H}$, and the electromagnetic fields satisfy the schrodinger's equation with this hamiltonian. Below, we provide expressions for the electric field operator $\hat{\textbf{E}}(\textbf{x})$ and the hamiltonian $\hat{H}$ for bulk media, lossless single-mode cavity and a single-mode waveguide (refer to Appendix~\ref{sec:em_quant} for a detailed derivation):
\begin{enumerate}
    \item \emph{Single mode cavity}: A lossless single-mode cavity can be described completely by the mode field profile, $\textbf{E}_C(\textbf{x})$ oscillating at its resonant frequency $\omega_C$. This mode has annihilation operator $a$ which satisfies the commutation $[\hat{a}, \hat{a}^\dagger] = 1$. The hamitlonian for the single mode cavity reduces to $H = \omega_C \hat{a}^\dagger \hat{a}$ and the electric field operator can be expressed as:
    \begin{align}
        \hat{\textbf{E}}(\textbf{x}) = \bigg[\frac{\hbar \omega_C}{2\varepsilon_0 \int_\Omega \varepsilon(\textbf{x})|\textbf{E}_C(\textbf{x})|^2 \textrm{d}^3\textbf{x}} \bigg]^{1/2} \textbf{E}_C(\textbf{x}) \hat{a} + \textrm{h.c.}
    \end{align}
    where $\Omega$ refers to the interior of the cavity.
    \item \emph{Single-mode waveguides}: The propagation of light within a single-mode waveguide can be described completely by a continuum of modes indexed by the wavenumber $\beta$, a dispersion relationship indicating the frequency $\omega(\beta)$ of the propagating modes and their mode profiles $\textbf{E}(\textbf{x}; \beta)$ --- assuming that the propagation axis is $x$ and $\boldsymbol{\rho}$ denotes the transverse coordinate, this mode has the form $\textbf{E}(\boldsymbol{\rho}; \beta) \exp(\textrm{i}\beta x)$. Each mode has an associated annihilation operator $\hat{a}(\beta)$, which satisfies the commutator $[\hat{a}(\beta), \hat{a}^\dagger(\beta')] = \delta(\beta - \beta')$ and $[\hat{a}(\beta), \hat{a}(\beta')] = 0$. The hamiltonian for the single-mode waveguide reduces to:
    \begin{align}
        H = \int_0^\infty \omega(\beta) \hat{a}^\dagger(\beta) \hat{a}(\beta) \textrm{d}\beta
    \end{align}
    with the electric field operator being given by:
    \begin{align}
        \hat{\textbf{E}}(\textbf{x}) = \int_0^\infty \bigg[\frac{\hbar \omega(\beta)}{2\varepsilon_0 \int_\Gamma \varepsilon(\boldsymbol{\rho})|\textbf{E}(\boldsymbol{\rho}; \beta)|^2 \textrm{d}^2\boldsymbol{\rho}} \bigg]^{1/2}\textbf{E}(\boldsymbol{\rho}; \beta) \exp(\textrm{i}\beta x)  \hat{a}(\beta) \textrm{d}\beta + \textrm{h.c.}
    \end{align}
    where we have assumed the waveguide mode to support propagation in only one direction and $\Gamma$ is the cross-section of the waveguide. Note that propagation in the backward direction can be modelled as another single-mode waveguide. Alternatively, defining the annihilation operator in terms of frequency $\omega$ instead of the propagation constant, $\hat{a}(\omega) = \hat{a}(\beta(\omega))/ \sqrt{v_G(\omega)}$ (where $v_G(\omega)$ is the group velocity of the waveguide mode at frequency $\omega$), which has commutation relations $[\hat{a}(\omega), \hat{a}^\dagger(\omega')] = \delta(\omega - \omega')$ and $[\hat{a}(\omega), \hat{a}(\omega')] = 0$, the hamiltonian can be expressed as:
    \begin{align}
        \hat{H} = \int_{\omega_c}^\infty \omega \hat{a}^\dagger(\omega) \hat{a}(\omega) \textrm{d}\omega
    \end{align}
    where $\omega_c$ is the cutoff frequency of the waveguide mode, and the electric field operator can be expressed as:
    \begin{align}
        \hat{\textbf{E}}(\textbf{x}) = \int_{\omega_c}^\infty \bigg[\frac{\hbar \omega}{2\varepsilon_0 v_G(\omega) \int_\Gamma \varepsilon(\boldsymbol{\rho}) |\textbf{E}(\boldsymbol{\rho}; \beta(\omega))|^2 \textrm{d}^2 \boldsymbol{\rho}} \bigg]^{1/2} \textbf{E}(\boldsymbol{\rho}; \omega)\exp[\textrm{i}\beta(\omega) x] \hat{a}(\omega) \textrm{d}\omega + \textrm{h.c.} 
    \end{align}
\end{enumerate}
Most quantum-optical systems that we will consider in this paper will comprise of a quantum system (such as a quantum dot, or coupled cavity-quantum dot system) being fed from an input (e.g. from a laser) and emitting into one or more single-mode waveguides. To understand the non-classical light generated by these sources, it is important to look more closely into the quantum states of light that can be supported in a waveguide mode. Almost all the quantum systems that we consider here emit light spectrally concentrated near a resonant frequency $\omega_0$ (e.g. transition frequency of a quantum dot, or resonant frequency of a cavity) --- in such a case, the dispersion relationship of the waveguide mode $\omega(\beta)$ can be linearly approximated around $\omega_0$, and the annihilation operator can be indexed by frequency $\omega$ instead of the propagation constant \cite{fan2010input} to obtain the following form of the Hamiltonian:
\begin{align}\label{eq:simple_hamiltonian_wg}
    H = \int_{-\infty}^{\infty} \omega \hat{a}^\dagger(\omega) \hat{a}(\omega) \textrm{d}\omega
\end{align}
where $[\hat{a}(\omega), \hat{a}^\dagger(\omega')] = \delta(\omega - \omega')$, $[a(\omega), a(\omega')] = 0$, and $a(\omega)$ [$a^\dagger(\omega)$] can be interpreted as an operator annihilating (creating) an excitation at frequency $\omega$. The electric field operator can also be simplified to:
\begin{align}
    \hat{\textbf{E}}(\textbf{x}) \approx \bigg[\frac{\hbar \omega_C}{2\varepsilon_0 v_G(\omega_C)} \bigg]^{1/2} \textbf{E}(\boldsymbol{\rho}; \beta(\omega_C)) \int_{-\infty}^{\infty} \hat{a}(\omega) \exp\bigg(\frac{\textrm{i}\omega x}{v_G}\bigg)\textrm{d}\omega + \textrm{h.c.}
\end{align}
This is equivalent to the `Markov' approximation in quantum optics --- it is equivalent to assuming that the waveguide is an inifinite bandwidth system. Note that the above expressions for the hamiltonian and the electric field operator also serve as good approximations for certain free-space modes e.g. describing the quantum properties of light propagating in a well-confined gaussian beam whose physics may otherwise be very complex to model \cite{shapiro2009quantum}. Throughout this paper, we will refer to modes (waveguide or free-space) that can be described with this simple model as loss-channels.

A single photon propagating in a bosonic loss-channel can be expressed as the following state:
\begin{align}\label{eq:single_photon_state}
    \ket{\Psi^{(1)}} = \int_{-\infty}^{\infty} \psi^{(1)}(\omega) \hat{a}^\dagger(\omega) \ket{\text{vac}} \textrm{d}\omega
\end{align}
where $\psi^{(1)}(\omega)$ is the (complex) amplitude of the contribution of $\omega$ to the single-photon state, and will satisfy the normalizing condition:
\begin{align}
    \int |\psi^{(1)}(\omega)|^2 \textrm{d}\omega = 1
\end{align}
In general, an $N$-photon state in the loss channel has the form:
\begin{align}\label{eq:N_photon_state}
    \ket{\Psi^{(N)}} &= \frac{1}{\sqrt{N !}}\int_{-\infty}^{\infty}\int_{-\infty}^{\infty} \dots \int_{-\infty}^{\infty} \psi^{(N)}(\omega_1, \omega_2 \dots \omega_N) 
    \prod_{i=1}^N \hat{a}^\dagger(\omega_i) \textrm{d}\omega_i \ket{\textrm{vac}}
\end{align}
where $\psi^{(N)}(\omega_1, \omega_2 \dots \omega_N)$ is symmetric with respect to permutation of the frequency arguments, and satisfies the normalization:
\begin{align}
    \int_{-\infty}^{\infty}\int_{-\infty}^{\infty} \dots \int_{-\infty}^{\infty} |\psi^{(N)}(\omega_1, \omega_2 \dots \omega_N) |^2 \textrm{d}\omega_1 \textrm{d}\omega_2 \dots \textrm{d}\omega_N = 1
\end{align}
An alternative way of expressing the state of the light in the loss channel is by using the `position' annihilation operator $\hat{a}(x)$ which is defined as the inverse fourier transform of the operator $\hat{a}(\omega)$:
\begin{align}
    \hat{a}(x) = \int \hat{a}(\omega) \exp\bigg(\frac{\textrm{i}\omega x}{v_G}\bigg) \frac{\textrm{d}\omega}{\sqrt{2\pi v_G}}
\end{align}
$a(x)$ [$a^\dagger(x)$] can be considered to annihilate (create) a photon at a position $x$ within the waveguide. The commutation relationship for $a(\omega)$ imply that $[a(x), a(x')] = 0$ and $[a(x), a^\dagger(x')] = \delta(x - x')$. The $N$-photon state as defined in Eq.~\ref{eq:N_photon_state} can be rewritten in terms of the position operator to obtain:
\begin{align}\label{eq:N_photon_state}
    \ket{\Psi^{(N)}} &= \frac{1}{\sqrt{N !}}\int_{-\infty}^{\infty}\int_{-\infty}^{\infty} \dots \int_{-\infty}^{\infty} \psi^{(N)}(x_1, x_2 \dots x_N) 
    \prod_{i=1}^N \hat{a}^\dagger(x_i) \textrm{d}x_i \ket{\textrm{vac}}
\end{align}
where $\psi^{(N)}(x_1, x_2 \dots x_N)$, the amplitude of finding the $N$ photons at $x_1, x_2 \dots x_N$, is related to $\psi^{(N)}(\omega_1, \omega_2 \dots \omega_N)$ via an inverse fourier transform:
\begin{align}
    \psi^{(N)}(x_1, x_2 \dots x_N) = \int_{-\infty}^{\infty} \int_{-\infty}^{\infty} \dots \int_{-\infty}^{\infty} \psi^{(N)}(\omega_1, \omega_2 \dots \omega_N) \prod_{i=1}^N \exp\bigg(-\frac{\textrm{i}\omega_i x_i}{v_G}\bigg) \frac{\textrm{d}\omega_i}{\sqrt{2\pi v_G}}
\end{align}
In the position basis representation for the $N$ photon state, its time evolution under the hamiltonian (Eq.~\ref{eq:simple_hamiltonian_wg}) is equivalent to translating the $\psi^{(N)}(x_1, x_2 \dots x_N)$ along every coordinate by the elapsed time. More explicitly:
\begin{align}
    \exp(-\textrm{i}H t) \ket{\Psi^{(N)}} = \frac{1}{\sqrt{N !}}\int_{-\infty}^{\infty}\int_{-\infty}^{\infty} \dots \int_{-\infty}^{\infty} \Psi^{(N)}(x_1 - v_G t, x_2 -v_G t \dots x_N-v_Gt) 
    \prod_{i=1}^N \hat{a}^\dagger(x_i) \textrm{d}x_i \ket{\textrm{vac}}
\end{align}
which follows directly from Eq.~\ref{eq:simple_hamiltonian_wg}, the definition of $a(x)$ and the commutators for $a(\omega)$ --- this intuitively corresponds to the light propagating down the loss channel with velocity $v_G$. 

The set of states $\{\ket{\Psi^{(k)}} \ \forall \ k \in \{0, 1, 2 \dots \} \}$ forms a complete basis for the hilbert space of the loss channel. The light emitted by quantum light sources into a loss channel can therefore be expressed as a superposition of such states:
\begin{align}
    \ket{\Psi} = \sqrt{P_0}\ket{\text{vac}} + \sum_{i = 1}^\infty \sqrt{P_i} \ket{\Psi^{(i)}}
\end{align}
where $P_i$ is the probability that the emitted light has $i$ photons ($P_0 + P_1 + P_2 +\dots = 1$). Of special interest are the states emitted by a single-photon source, and a coherent source which are described below.

A \emph{single-photon} source coupling completely to a single loss-channel would have emit a state given by:
\begin{align}\label{eq:ideal_single_ph_src_state}
    \ket{\Psi} = \sqrt{P_0}\ket{\text{vac}} + \sqrt{P_1} \ket{\Psi^{(1)}}
\end{align}
with $P_0 + P_1 = 1$. $P_1$ is the probability that a single-photon emission has happened, it is therefore desirable to make $P_1$ as large as possible. An important consideration here is that most single photon sources invariably couple to multiple loss channels --- these include not only the channels used for driving or taking an output from the source, but also loss channels that correspond to radiation losses that cannot be collected in a practical system. In such cases, the state of light in the output loss channel can no longer be described by a pure state but rather by a density matrix $\hat{\rho}$ --- assuming that the source only emits a single photon, $\hat{\rho}$ has the form:
\begin{align}\label{eq:single_ph_density_mat}
    \hat{\rho} = P_0\ket{\text{vac}}\bra{\text{vac}} + P_1\hat{\rho}^{(1)} + \hat{\rho}^{(\chi)}
\end{align}
where $P_0 + P_1 = 1$ and
\begin{align}
    \hat{\rho}^{(\chi)} &= \int_{-\infty}^\infty \big[ \rho^{(\chi)}(\omega) \hat{a}^\dagger(\omega) \ket{\text{vac}}\bra{\text{vac}} + \rho^{(\chi)*}(\omega) \ket{\text{vac}} \bra{\text{vac}} \hat{a}(\omega) \big] \textrm{d}\omega \\
    \hat{\rho}^{(1)} &= \int_{-\infty}^{\infty} \int_{-\infty}^\infty \rho^{(1)}(\omega_1, \omega_2) \hat{a}^\dagger(\omega_1) \ket{\text{vac}}\bra{\text{vac}}\hat{a}(\omega_2)\textrm{d}\omega_1 \textrm{d}\omega_2
\end{align}
where $\rho^{(1)}(\omega_1, \omega_2) = \rho^{(1)*}(\omega_2, \omega_1)$ and the following normalization condition is satisfied:
\begin{align}
    \text{Trace}(\hat{\rho}^{(1)}) = \int_{-\infty}^{\infty} \rho^{(1)}(\omega, \omega) \textrm{d}\omega = 1
\end{align}
Note that if the emitted light is in a pure state (i.e.~the source couples only to the output loss channel), then density matrix factorizes into a product of a bra and a ket: $\hat{\rho} = \ket{\Psi}\bra{\Psi}$. In terms of $\hat{\rho}^{(\chi)}$ and $\hat{\rho}^{(1)}$, this is equivalent to $\rho^{(\chi)}(\omega) = \sqrt{P_0 P_1} \psi^{(1)}(\omega)$ and $\rho^{(1)}(\omega_1, \omega_2) = \psi^{(1)}(\omega_1) \psi^{(1)*}(\omega_2)$. For a more general mixed state, it is possible to express the density matrix into a sum of such factorizable terms:
\begin{align}
    \hat{\rho} = \sum_{i} \sigma_i \ket{\Psi_i}\bra{\Psi_i}
\end{align}
where $1 \geq \sigma_1 \geq \sigma_2 \dots  \geq 0$, $\sigma_1 + \sigma_2 \dots = 1$ and $\bra{\Psi_i}\Psi_j\rangle = \delta_{i, j}$. This is referred to as the Schmidth decomposition \cite{lamata2005dealing} of the density matrix --- physically, this decomposition can be interpreted as indicating that the state is in a mixture of pure states $\ket{\Psi_i}$, with the contribution of the $i^\text{th}$ pure state being $\sigma_i$. The rank of this decomposition (i.e.~number of non-zeros or large $\sigma_i$) can be taken as an indicator of how `pure' the state is.

Below, we briefly define some metrics that can be used to characterize such a single-photon source:
\begin{enumerate}
\item \emph{Spectral content}: The spectral content $n_\omega$ of a single photon source can be defined by the average number of photons per unit frequency emitted by the source at a given frequency. This can be computed by averaging the operator $a_\omega^\dagger a_\omega$ over the state of the emitted light:
    \begin{align}
        n_\omega = \text{Trace}(\hat{\rho} a_\omega^\dagger a_\omega) = \text{Trace}(a_\omega \hat{\rho} a_\omega^\dagger) = P_1 \rho^{(1)}(\omega, \omega)
    \end{align}
    For a source emitting a pure state, this further simplifies to $n_\omega = P_1 |\psi^{(1)}(\omega)|^2$.
    \item \emph{Brightness}: The brightness of the source can be defined as the average number of photons emitted by the source i.e. integrating $n_\omega$ over all frequencies --- this evaluates to $P_1$, which is also equal to the probability of the source emitting a single photon.
    \item \emph{Trace purity}: The trace purity of the single photon source characterizes how close to a pure state is the emitted single photon state. Mathematically, it is defined by:
    \begin{align}
        \text{Trace}(\hat{\rho}^2) = P_0^2 + 2 \bigg | \int_{-\infty}^{\infty} \rho^{(\chi)}(\omega)\textrm{d}\omega \bigg |^2 + P_1^2 \int_{-\infty}^\infty \rho^{(1)}(\omega, \omega) \textrm{d}\omega
    \end{align}
    For the emitted light being in a pure state, it can be readily verified that the trace purity evaluates to 1. A mixed state would have a trace purity smaller than 1, with its deviation from 1 signifying the extent to which it is not a pure state.
\end{enumerate}

A \emph{coherent state} is the state of light emitted from a laser source:
\begin{align}\label{eq:coherent_state}
    \ket{\alpha(\omega)} = \exp\bigg(\int \big[\alpha^*(\omega)\hat{a}(\omega) - \alpha(\omega)\hat{a}^\dagger(\omega)\big] \textrm{d}\omega\bigg) \ket{\text{vac}}
\end{align}
where $\alpha(\omega)$ is the complex spectrum of the coherent state. This state is a common eigenvector of all annihilation operator $\hat{a}(\omega)$, satisfying the eigenvalue equation $\hat{a}(\omega) \ket{\alpha(\omega)} = \alpha(\omega) \ket{\alpha(\omega)}$. The mean photon number in such a coherent state is given by $\int |\alpha(\omega)|^2 \textrm{d}\omega$, and can be taken as a measure of the brightness of the state. Note that unlike the state emitted by a single-photon source, the mean photon number can be arbitrarily large if sufficient energy is put into the coherent state. The coherent state is regarded to be the closest approximation of a classical state of light within all possible quantum states \cite{loudon2000quantum}.

A very popular experimental metric for characterizing these states is the two-time second-order correlation function $g^{(2)}(t_1, t_2)$, which is a measure of the joint probability of detecting a photon at both the time instants $t_1$ and $t_2$ \cite{mandel1995optical}. Mathematically, assuming that the detector is located at $x = L$, this is defined by (note that all the operators are the position annihilation operators):
\begin{align}\label{eq:two_time_corr}
    g^{(2)}(t_1, t_2) &= \frac{\bra{\Psi}\hat{a}^\dagger(t_1; L) \hat{a}^\dagger(t_2; L) \hat{a}(t_2; L)\hat{a}(t_1; L) \ket{\Psi}}{\bra{\Psi} \hat{a}^\dagger(t_1; L) \hat{a}(t_1; L) \ket{\Psi}\bra{\Psi} \hat{a}^\dagger(t_2; L) \hat{a}(t_2; L) \ket{\Psi}} \nonumber \\
    &=\frac{\bra{\Psi}\hat{a}^\dagger(L-v_G t_1) \hat{a}^\dagger(L-v_G t_2) \hat{a}(L-v_G t_2)\hat{a}(L-v_G t_1) \ket{\Psi}}{\bra{\Psi} \hat{a}^\dagger(L-v_Gt_1) \hat{a}(L-v_Gt_1) \ket{\Psi}\bra{\Psi} \hat{a}^\dagger(L-v_G t_2) \hat{a}(L-v_G t_2) \ket{\Psi}}
\end{align}
Where $a(t; L) = a(L-v_G t)$ is the position annihilation operator at the location of the detector in the heisenberg picture at time $t$. In a realistic experimental setting, however, the photodetector usually has a response time which is much larger than the temporal width of the light being emitted from the light source. This practical limitation prohibits the exact measurement of $g^{(2)}(t_1, t_2)$ as a function of $(t_1, t_2)$, rather it allows the measurement of the integrated two-time correlation $g^{(2)}[0]$ \cite{Fischer2016}:
\begin{align}
    g^{(2)}[0] = \frac{\int \int \bra{\Psi}\hat{a}^\dagger(t_1; L) \hat{a}^\dagger(t_2; L) \hat{a}(t_2; L)\hat{a}(t_1; L) \ket{\Psi}\textrm{d}t_1 \textrm{d}t_2}{[\int \bra{\Psi} \hat{a}^\dagger(t; L) \hat{a}(t; L) \ket{\Psi}\textrm{d}t]^2}
\end{align}
Both $g^{(2)}(t_1, t_2)$ and $g^{(2)}[0]$ have remarkably different values for different states of light, making it a suitable tool for characterizing the state of light emitted by a light source. In particular,
\begin{enumerate}
    \item For the light emitted by an ideal single-photon source (Eq.~\ref{eq:ideal_single_ph_src_state}), $g^{(2)}(t_1, t_2) = 0$ and $g^{(2)}[0] = 0$ --- a small value of $g^{(2)}[0]$ is often treated is an experimental signature of a single-photon state.
    \item For light emitted from an ideal coherent source (Eq.~\ref{eq:coherent_state}), $g^{(2)}(t_1, t_2) = 1$ and $g^{(2)}[0] = 1$.
    \item For a two-photon state (Eq.~\ref{eq:N_photon_state} with $N = 2$),
    \begin{align}
        g^{(2)}(t_1, t_2) = \frac{|\psi^{(2)}(x_1 = L - v_G t_1, x_2 = L - v_G t_2)|^2}{2\int |\psi^{(2)}(x_1 = L - v_G t_1, x_2)|^2 \textrm{d}x_2\int |\psi^{(2)}(x_1, x_2 = L - v_G t_2)|^2 \textrm{d}x_1} \leq \frac{1}{2}
    \end{align}
    and $g^{(2)}[0] = 1/2$.
\end{enumerate}
Similarly, it is possible to define the two time first order correlation function, $g^{(1)}(t_1, t_2)$, via:
\begin{align}
    g^{(1)}(t_1, t_2) = \frac{\bra{\Psi}\hat{a}^\dagger(t_1; L) \hat{a}(t_2; L)\ket{\Psi}}{\sqrt{\bra{\Psi}\hat{a}^\dagger(t_1; L) \hat{a}(t_1; L)\ket{\Psi}\bra{\Psi}\hat{a}^\dagger(t_2; L) \hat{a}(t_2; L)\ket{\Psi}}}
\end{align}
and its integrated version:
\begin{align}
    g^{(1)}[0] = \frac{\int \int \bra{\Psi}\hat{a}^\dagger(t_1; L) \hat{a}(t_2; L)\ket{\Psi}\textrm{d}t_1\textrm{d}t_2}{\int\bra{\Psi}\hat{a}(t; L)^\dagger(t)\hat{a}(t; L) \ket{\Psi}\textrm{d}t}
\end{align}
This first order correlation function is not an experimentally measurable quantity, but can be used as a theoretical tool to gain information about the amplitude and phase spectrum of the light propagating in the loss-channel (note that it is a complex number, and carries information of the phase relationship between the light incident on the detector at $t=t_1$ and $t=t_2$).
\begin{figure*}[!t]
  \includegraphics[scale=0.8]{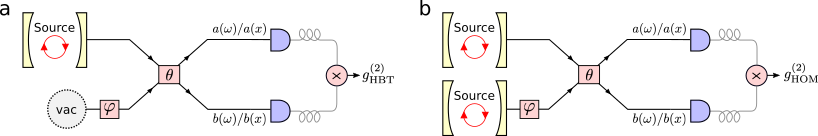}
  \caption{\textbf{Interferometers for characterizing light sources:} Schematic of \textbf{a} Hanbury Brown Twiss interferometer \textbf{b} Hong-Ou Mandel interferometer. In both cases, the beam splitter is assumed to be a 50-50 beam splitter ($\theta = \pi / 4$) and the phase shift $\phi$ models random variations between the path lengths of the two branches of the interferometer.  (Figure adapted from \cite{Fischer2016})}
  \label{figure:interferometer}
\end{figure*}

An experimental difficulty in the measurement of the two-time second-order correlation function is that most photo-detectors have a dead time --- this refers to the time after a photo-detection for which the detector is no longer able to register incoming photons. This dead time is often larger than the pulse-width of the light emitted by the photon source, thereby making it impossible to measure the two-time correlation with just one detector. An alternative scheme to get around this issue is the Hanbury-Brown and Twiss interferometer (HBT) --- this scheme (Fig.~\ref{figure:interferometer}\textbf{a}) splits the light emitted by the source into two waveguides via a 50-50 beam splitter ($\theta = \pi / 4$), and then performs photodetection in the two waveguides separately \cite{Fischer2016}. The phase shift $\varphi$ shown in the setup models random variation in the path lengths of the two arms of the interferometer due to environmental fluctations --- the result of an experimental measurement performed over repeated trial effectively averages over this phase shift. The two-time correlation function measured by this setup is then given by:
\begin{align}
    g_\text{HBT}^{(2)}(t_1, t_2) \approx \frac{\mathbb{E}_\varphi[\bra{\Psi} \hat{a}^\dagger(t_1; L) \hat{b}^\dagger(t_2; L) \hat{b}(t_2; L) \hat{a}(t_1; L) \ket{\Psi}]}{\mathbb{E}_\varphi[\bra{\Psi}\hat{a}^\dagger(t_1; L) \hat{a}(t_1; L) \ket{\Psi}] \mathbb{E}_\varphi[ \bra{\Psi}\hat{b}^\dagger(t_2; L) \hat{b}(t_2; L) \ket{\Psi}]}
\end{align}
and it's integrated version:
\begin{align}
    g_\text{HBT}^{(2)}[0] \approx \frac{\int \int \mathbb{E}_\varphi[\bra{\Psi} \hat{a}^\dagger(t_1; L) \hat{b}^\dagger(t_2; L) \hat{b}^\dagger(t_2; L) \hat{a}^\dagger(t_1;L) \ket{\Psi}\textrm{d}t_1 \textrm{d}t_2]}{\int \mathbb{E}_\varphi[\bra{\Psi}\hat{a}^\dagger(t; L) \hat{a}(t; L)  
     \ket{\Psi}\textrm{d}t] \int \mathbb{E}_\varphi[\bra{\Psi}\hat{b}^\dagger(t; L) \hat{b}(t; L) \ket{\Psi}\textrm{d}t]}
\end{align}
As is shown in appendix, this is identical to the two-time correlation function defined in Eq.~\ref{eq:two_time_corr}.
For single-photon sources, another important performance metric is the indistinguishability of the single-photon sources --- the fidelity of any information processing or computing scheme that relies on the ability to entangle the light emitted by two single-photon sources would ultimately be limited by how distinguishable the single-photon sources are. Experimentally, the indstinguishability of two single-photon sources can be determined by the Hong-Ou Mandel interferometer (HOM) --- this scheme (Fig.~\ref{figure:interferometer}\textbf{b}) combines the two single-photons at a beam splitter and performs photodetection at the two waveguides separately \cite{Fischer2016}. We define the integrated two-time correlation for this interferometer $g^{(2)}_\text{HOM}[0]$ via:
\begin{align}
    g^{(2)}_\text{HOM}[0] \approx \frac{\int \int  \mathbb{E}_\varphi[\bra{\Psi} \hat{a}(t_1; L) \hat{b}(t_2; L) \hat{b}^\dagger(t_2; L) \hat{a}^\dagger(t_1; L) \ket{\Psi}] \textrm{d}t_1 \textrm{d}t_2}{\int \mathbb{E}_\varphi[\bra{\Psi}\hat{a}^\dagger(t; L) \hat{a}(t; L) \ket{\Psi}]\textrm{d}t \int \mathbb{E}_\varphi[\bra{\Psi}\hat{b}^\dagger(t; L) \hat{b}(t; L) \ket{\Psi}] \textrm{d}t}
\end{align}
In case both the sources are assumed to be identical, using Eqs.~\ref{eq:decomposition_two_time} and \ref{eq:decomposition_one_time} from Appendix \ref{sec:interferometers}, $g^{(2)}_\text{HOM}[0]$ can be expressed in terms of the visibility $v$ and the integrated second-order two-time correlation $g^{(2)}[0]$ of the source:
\begin{align}
    g^{(2)}_\text{HOM}[0] = \frac{g^{(2)}[0]}{2} + \frac{1 - v}{2}
\end{align}
where $v$ is given by:
\begin{align}
    v = \frac{\int \int |\bra{\Psi} a^\dagger(t_1; L)a(t_2; L \ket{\Psi}|^2\textrm{d}t_1 \textrm{d}t_2}{[\int \bra{\Psi}a^\dagger(t; L)a(t; L) \ket{\Psi}\textrm{d}t]}
\end{align}
The parameter $v$ is correlated with how pure the state emitted by the single-photon source ($v = 1$ is indicative of a pure state) is and $g^{(2)}[0]$ is a measure of how significant the multi-photon contributions to the emission from the source are \cite{Fischer2018}. This decomposition thus decomposes  $g^{(2)}_\text{HOM}[0]$ into two contributing factors.

For example, as is shown in the appendix, for a Hong-Ou-Mandel interference between two single-photon sources, this correlation function can be expressed in terms of the density matrices of the two sources:
\begin{align}
    g^{(2)}_\text{HOM}[0] = \frac{2 P_{1, a} P_{2, a}}{(P_{1, a} + P_{2, a})^2}\bigg[1 - \int_{-\infty}^{\infty}\int_{-\infty}^\infty \rho_a^{(1)*}(\omega_1, \omega_2) \rho_b^{(1)}(\omega_1, \omega_2) \textrm{d}\omega_1 \textrm{d}\omega_2 \bigg]
\end{align}
where the subscripts $a$ and $b$ refer to whether the sources emit into the waveguide labeled as $a$ or $a$, and $P_1$ and $\rho^{(1)}(\omega_1, \omega_2)$ have been defined in Eq.~\ref{eq:single_ph_density_mat}. Measurement of $g^{(2)}_\text{HOM}[0]$ provides information about the purity and indistinguishability of the two photons. To see this, it is useful to specialize this expression for two limiting cases:
\begin{enumerate}
    \item \emph{Both sources emitting pure states}: In this case, $\rho^{(1)}(\omega_1, \omega_2) = \psi^{(1)*}(\omega_1)\psi^{(1)}(\omega_2)$ (where $\Psi^{(1)}(\omega)$ is defined in Eq.~\ref{eq:ideal_single_ph_src_state}), and $g^{(2)}_\text{HOM}[0]$ reduces to:
    \begin{align}
        g^{(2)}_\text{HOM}[0] = \frac{2P_{1,a}P_{1, b}}{(P_{1, a} + P_{1, b})^2}\bigg[1 - \bigg|\int_{-\infty}^\infty \psi^{(1)*}_a(\omega) \psi^{(1)}_b(\omega) \textrm{d}\omega \bigg|^2 \bigg]
    \end{align}
    from which it can clearly be seen that $g^{(2)}_\text{HOM}[0] = 0$ indicates that the two sources have identical complex spectras. $g^{(2)}_\text{HOM}[0]$ thus proovides information about how distinguishable two single photon sources are.
    \item \emph{Both sources being identical}: In the case where both sources are identical and emit single photons with unity probability, but not necessarily in a pure state, $g^{(2)}_\text{HOM}[0]$ simplifies to:
    \begin{align}
        g^{(2)}_\text{HOM}[0] = \frac{1}{2}\bigg[1 - \int_{-\infty}^{\infty} \int_{-\infty}^{\infty}|\rho^{(1)}(\omega_1, \omega_2)|^2 \textrm{d}\omega_1 \textrm{d}\omega_2 \bigg] = \frac{1}{2} \big[1 - \text{Tr}(\hat{\rho}^2)\big]
    \end{align}
    and the visibility $v = \text{Tr}(\hat{\rho}^2)$. In this case, it can be seen that $g^{(2)}_\text{HOM}[0]$ (or visibility) provides direct information about the trace purity of the single-photon sources --- a pure state has $g^{(2)}_\text{HOM}[0] = 0$.  While the trace purity is not in general accessible from a Hong--Ou--Mandel experiment when there are multiphoton components of emission present for the sources, recent work has shown how to access the trace purity using methods beyond the scope of this review \cite{Fischer2018}. 
\end{enumerate}

% %
% \begin{figure*}[!t]
%   \includegraphics[width=12cm]{Figure-2b3.pdf}
%   \caption{\textbf{MZ setup} Schematic illustration of an unbalanced Mach-Zehnder interfeormeter which is used to measure the indistinguishability of photons emitted from a single source. (Figure from \cite{Fischer2016})}
%   \label{figure:2b3}
% \end{figure*}
%
\subsection{Semiconductor quantum dots}
\label{subsection:QD}
\begin{figure}[!t]
  \includegraphics[width=8.6cm]{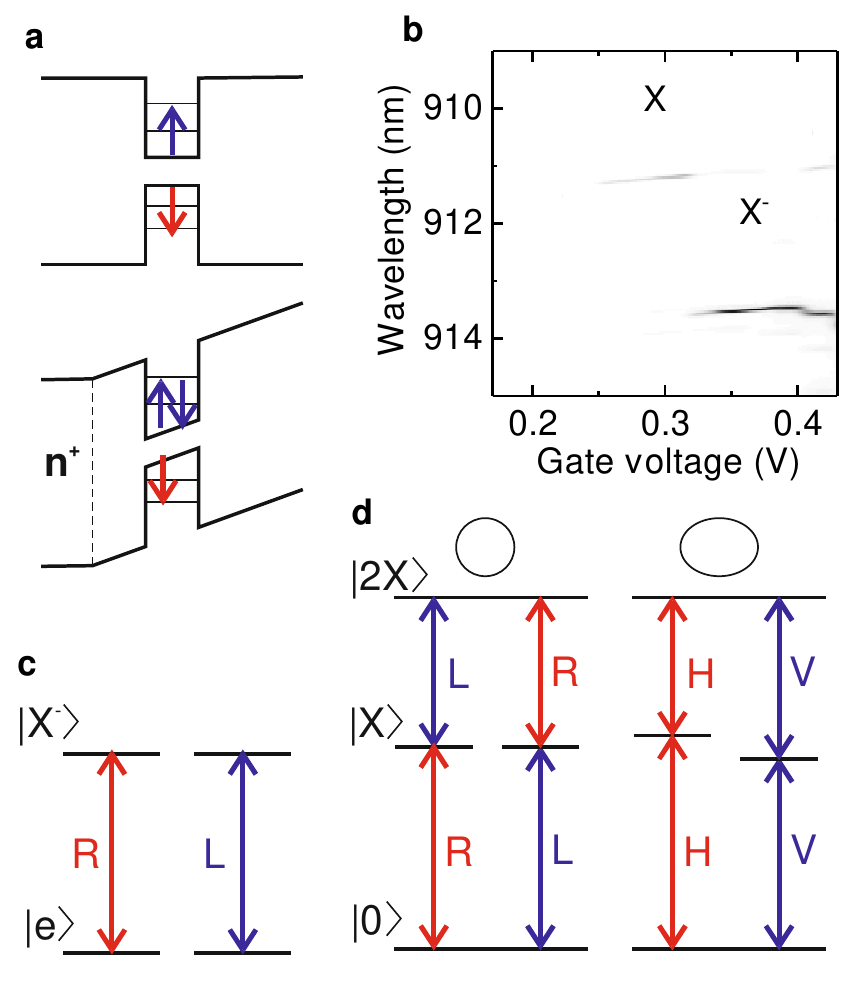}
  \caption{\textbf{Semiconductor quantum dots: a} Schematic illustration of band structure in growth direction (top) without and (bottom) with applied electric field. \textbf{b} Gate voltage dependent photoluminescence measurement illustrating the DC-Stark effect and sequential charging. \textbf{c} Level scheme with optical transitions of a singly-charged QD. \textbf{d} Level scheme with optical transitions of an uncharged QD with (left) and without (right) $D_{2D}$ symmetry. (b adopted from \cite{Hanschke2018})}
  \label{figure:2a1}
\end{figure}
QDs are nanometer-sized islands of a lower bandgap material embedded in a matrix of larger bandgap material \cite{Arakawa1982}. This leads to a three-dimensional confinement for electrons and holes with discrete shell-like energy levels which exhibit strong optical interband transitions (figure \ref{figure:2a1}a). These transitions have a near-unity quantum efficiency and almost exclusive emission into the ZPL which makes QDs ideal for sources of non-classical light. The energies of these transitions results from the bandgap, confinement energy, strain and Coulomb interactions between the charge carriers.

Typically, QDs are grown by molecular beam epitaxy (MBE) or metalorganic chemical vapour deposition (MOCVD) and over the years different material combinations and growth techniques have been established. The most common QDs are InAs/GaAs QDs grown by strain-driven self-assembly (Stranski-Krastanov (SK) growth) where the lattice mismatch between InAs and GaAs leads to the formation of QDs after a certain critical thickness of InAs is reached. Depending on the growth conditions, In(Ga)As/GaAs QDs typically emit in the range 900-1300 nm. To obtain InAs QDs emitting in the telecommunication C-band (1530-1565 nm) the lattice mismatch has to be reduced which can be achieved by using InP or AlGaInAs as the substrate \cite{Enzmann2010, Benyoucef2013, Miyazawa2016}. In addition, it was recently demonstrated that metamorphic InGaAs strain-relaxation layers can also be used to shifts the emission of InAs QDs into the C-band \cite{Paul2017}. In addition to the SK growth, droplet epitaxy (DE) of QDs has been established which results in strain-free QDs. Here, demonstrated material combinations are GaAs/AlGaAs QDs emitting at 780nm \cite{Heyn2009, Sonnenberg2012, Heyn2017}, InAs/GaAs QDs emitting around 900nm  \cite{Skiba2017} and InAs/InP QDs emitting at 1550nm \cite{Skiba2017}. A different method for growing QDs are nanowire-QDs where during the 1D growth of InP nanowires a InAsP segment is grown which forms a QD \cite{Reimer2012}. 

Embedding QDs into diode structures (figure \ref{figure:2a1}a) allows to tune their emission energies via the DC-Stark effect. Moreover, by carefully choosing the distance between QDs and an n-doped or p-doped reservoir allows to deterministically charge the QDs with electrons or holes \cite{Warburton2000}. If the lowest electron level in the QD is lowered below the Fermi energy the QD is deterministically charged with a single electron. Figure \ref{figure:2a1}b shows a gate-voltage dependent photoluminescence measurement of a QD under non-resonant excitation and reveals clear stability plateaus which can be assigned to the charge neutral exciton X which consists of of one electron and one hole and a singly-charged exciton X$^-$ which consists of two electrons and one hole. In addition, embedding  QDs into diode structures also stabilize the electronic environment which reduces inhomogeneous broadening and allows for nearly transform-limited linewidth \cite{Matthiesen2012, Kuhlmann2015}. To these ends, also mechanisms for active stabilization have been demonstrated \cite{Prechtel2013, Hansom2014}.

Due to the confinement and strain, the lowest energy levels for holes have an angular momentum of $J_z=\pm\frac{3}{2}$ while electrons have a spin of $S_z=\pm\frac{1}{2}$. Therefore, each orbital state has a spin degeneracy of 2 and singly-charged QDs exhibit two optical transitions which couple to right-handed and left-handed circularly polarized light (Fig.~\ref{figure:2a1}\textbf{c}). The s-shell of uncharged QDs can host two-electrons and two-holes which is called the bi-exciton state and labelled 2X or XX (Fig.~\ref{figure:2a1}\textbf{d}). Due to the Coulomb interaction the transition X-2X has a smaller energy than the 0-X transition and the energy difference is often called binding energy. For perfectly symmetric QDs the two X states are degenerate and couple to 0 and 2X via circularly polarized light. However, in the presence of any deviation from this symmetry due to the exchange interaction the degeneracy of the X states is lifted, they are separated by a finestructure splitting (FSS - typically $< 100 \mu$eV) and couple to linearly polarized light \cite{Bayer2002}.

As a consequence, only singly-charged QDs can be treated as true two-level systems. When exciting a neutral QD with a narrow-band laser the two X states can be addressed individually and treated as two-level systems. However, when exciting a neutral QD with short pulses which are spectrally broader than the FSS where the polarization is different than H or V a coherent superposition of the two X states is created which precesses in time. This is particularly important for resonant spectroscopy where the excitation laser is typically suppressed from the QDs emission by using orthogonal linear polarizations for excitation and detection. For a charged QD excitation and detection channel couple to the same transition though with a reduced strength due to the polarization mismatch. However, when exciting an uncharged QD with a short optical laser pulse the observed signal strongly depends on the choice of polarization relative to the QDs axis \cite{Benny2011}. H or V polarized light will excite the QD in one of the two X states and since the emission has the same polarization it will be suppressed in the detection channel. On the other hand, exciting with a diagonal polarization will create a precessing superposition such that the orthogonally polarized detection channel will collect signal which oscillates in time \cite{dada2016indistinguishable}.

\subsection{Light-matter interaction}
Light-matter interactions lie at the heart of non-classical light generation --- a full quantum treatment of the emitter and the optical field is required to model such interactions at the quantum level). Fundamentally, the interaction of an emitter interacts with an electromagnetic field is due to its dipole moment --- in the language of quantum mechanics, this dipole is described by an operator $\hat{\boldsymbol{\mu}} = -e\hat{\textbf{x}}$ where $\hat{\textbf{x}}$ is the operator corresponding to the difference between the positions of positive and negative charge in the emitter. Denoting by $\ket{g}$ and $\ket{e}$ as the two states between which an optical transition happens, this operator can be expressed as:
\begin{align}
    \hat{\boldsymbol{\mu}}= \boldsymbol{\mu} \hat{\sigma} + \boldsymbol{\mu}^*\sigma^\dagger
\end{align}
where $\sigma = \ket{g}\bra{e}$, $\sigma^\dagger = \ket{e}\bra{g}$ and we has assumed that $\bra{e} \hat{\textbf{x}} \ket{e} = \bra{g} \hat{\textbf{x}} \ket{g} = 0$ (this is true if the states of the emitter's electron has symmetry or anti-symmetry with respect to reflection across the origin, which is usually true for most solid state emitters of interest). Since the variance in the position of electrons in the emitter is typically much smaller than the wavelength of light, the interaction hamiltonian between the emitter and electromagnetic field can be approximated by \cite{cohen1998atom}:
\begin{align}
    \hat{H}_I = -\hat{\boldsymbol{\mu}}\cdot \hat{\textbf{E}}(\textbf{x}_D)
\end{align}
where $\hat{\textbf{E}}(\textbf{x}_D)$ is the electric field operator at the (mean) position $\textbf{x}_D$ of the emitter's dipole. The dynamics of an emitter emerging from such an interaction hamiltonian can have vastly different characteristics depending on the electromagnetic structure. For e.g. using this hamiltonian to study the dynamics of an emitter coupling to bulk homogenous media, the phenomenon of spontaneous emission and lamb shifts in the emitter frequency have been successfully explained \cite{cohen1998atom}. Below, we briefly review the basic dynamics of emitters coupling to optical cavities and waveguides.

\emph{Emitters coupling to cavities}: In this case, the interaction hamiltonian simplifies to:
\begin{align}
    \hat{H}_I = g \sigma^\dagger a + f \sigma a + \text{h.c.}
\end{align}
where
\begin{align}
    g =  -\bigg[\frac{\hbar \omega_C}{2\varepsilon_0 \int_\Omega \varepsilon(\textbf{x})|\textbf{E}_C(\textbf{x})|^2 \textrm{d}^3\textbf{x}} \bigg]^{1/2} \boldsymbol{\mu}^*\cdot\textbf{E}_C(\textbf{x}_D) \text{ and } f = -\bigg[\frac{\hbar \omega_C}{2\varepsilon_0 \int_\Omega \varepsilon(\textbf{x})|\textbf{E}_C(\textbf{x})|^2 \textrm{d}^3\textbf{x}} \bigg]^{1/2} \boldsymbol{\mu}\cdot\textbf{E}_C(\textbf{x}_D)
\end{align}
Note that the interaction hamiltonian $\hat{H}_I$ has terms that correspond to the simulataneous creation (annihilation) of a photon and excitation (decay) of the emitter to its excited state (ground state) --- $\hat{a}^\dagger \hat{\sigma}^\dagger$ ($\hat{a}\hat{\sigma}$). This term produces extremely high-frequency contribution ($\sim \omega_C + \omega_E$, where $\omega_E$ is the frequency corresponding to the transition energy of the emitter) to the evolution of the system --- if $|f|, |g| \ll \omega_C,  \omega_E$, then the contribution of such terms to the interaction hamiltonian can be ignored:
\begin{align}
    \hat{H}_I \approx g \sigma a^\dagger + g^* a \sigma^\dagger
\end{align}
This approximation is often called the rotating wave approximation --- the resulting hamiltonian, called the Jaynes-Cumming interaction hamiltonian, lies at the heart of cavity quantum electrodynamics. The coupling constant $g$, which governs the strength of the cavity-emitter interaction, depends on the position of the dipole and how well the cavity mode is confined. By defining the mode volume $V_M$ of the cavity by:
\begin{align}
    V_M = \frac{\int_\Omega \varepsilon(\textbf{x}) | \textbf{E}_C(\textbf{x})|^2 \textrm{d}^3\textbf{x}}{\max[\varepsilon(\textbf{x}) |\textbf{E}_C(\textbf{x})|^2]}
\end{align}
This coupling constant can be re-expressed as:
\begin{align}
    g = \bigg[\frac{\hbar \omega_C}{2\varepsilon_0 V_M} \bigg]^{1/2} \mathcal{F}(\textbf{x}_D) \ \text{where } \mathcal{F}(\textbf{x}_D) = -\frac{\boldsymbol{\mu}^* \cdot \textbf{E}_C(\textbf{x}_D)}{\sqrt{\max[\varepsilon(\textbf{x}) |\textbf{E}_C(\textbf{x})|^2]}}
\end{align}
$\mathcal{F}(\textbf{x}_D)$ is a dimensionless factor which governs the overlap between the cavity mode and the emitter --- this can be increased by placing the emitter at the field maxima, and aligning it's dipole with the field polarization. Clearly, decreasing the mode volume $V_M$ increases the strength of light-matter interaction --- an overview of the mode volumes achievable by typical optical cavity designs is given in section. The full hamiltonian for the cavity-emitter system, including the energies of the emitter and the cavity:
\begin{align}
    \hat{H} = \omega_C \hat{a}^\dagger \hat{a} + \omega_E \hat{\sigma}^\dagger \hat{\sigma} + \hat{H}_I
\end{align}
wherein we have expressed the emitter hamiltonian (assuming the ground state to be at 0 energy) as $\omega_E \ket{e}\bra{e} = \omega_E \sigma^\dagger \sigma$. As a consequence of the rotating wave approximation, this hamiltonian conserves the total excitation number $\hat{n} = \hat{a}^\dagger\hat{a} + \hat{\sigma}^\dagger \hat{\sigma}$  for the emitter-cavity system (this can be shown by verifying that $[\hat{H}, \hat{n}] = 0$). This hamiltonian can be exactly diagonalized --- it is straightforward to show that there are two $n$ excitation eigenstates of this hamiltonian (which are entangled states of the cavity mode and the emitter, often called \emph{polaritons}) with eigen-frequencies $\omega_n^{(+)}$ and $\omega_n^{(-)}$ given by:
\begin{align}
    \omega_n^{(\pm)} = \frac{(2n - 1) \omega_C + \omega_E}{2} \pm \bigg[n |g|^2 + \frac{(\omega_C - \omega_E)^2}{4} \bigg]^{1/2} 
\end{align}
These eigen-frequencies are plotted in Fig.~\ref{figure:3c1}\textbf{a} as a function of the emitter frequency --- the development of the gap that develops between the two frequency curves due to the coupling between the emitter and cavity is refered to as the normal mode splitting. A consequence of such a splitting between is a coherent exchange of energy between the emitter and cavity if one of them is initially excited. For e.g.~if the emitter is initialized in its excited state $\ket{\Psi(t = 0)} = \ket{e} \ket{\text{vac}}$, then the probability $P_e(t)$ of the emitter being in its excited state at time $t$ is given by:
\begin{align}
    P_e(t) = 1 - \frac{|g|^2}{4\Gamma^2}\sin^2(\Gamma t)
\end{align}
where $\Gamma^2 = |g|^2 + (\omega_C - \omega_E)^2 / 4$. The cavity and emitter thus exchange energy periodically --- this is in stark contrast to an exponential decay of the emitter energy when they couple to a continuum of modes (e.g. bulk media or waveguide --- see below). For a lossy cavity, these oscillations are damped with the oscillatory behaviour being increasingly stronger with an increase in the quality factor of the cavity.

\emph{Emitters coupling to waveguides}: Within the rotating-wave and Markov approximation, we obtain the following interaction hamiltonian for the emitter-waveguide system:
\begin{align}
    \hat{H}_I = \int_{-\infty}^\infty \big[g \sigma a^\dagger(\omega) + g^* a(\omega) \sigma^\dagger \big] \textrm{d}\omega
\end{align}
where $g$ is given by:
\begin{align}
    g = -\bigg[\frac{\hbar \omega_E}{2\varepsilon_0 \int_\Gamma \varepsilon(\boldsymbol{\rho}) |E(\boldsymbol{\rho}; \beta(\omega_E))|^2 \textrm{d}^2 \boldsymbol{\rho}} \bigg]^{1/2} \boldsymbol{\mu}^* \cdot \textbf{E}(\boldsymbol{\rho}_E; \beta(\omega_E))
\end{align}
where we have assumed that the emitter is located at $x = 0$ with a transverse coordinate $\boldsymbol{\rho}_E$. Alternatively, this coupling constant can be expressed as:
\begin{align}
    g = -\bigg[\frac{\hbar \omega_E}{2\varepsilon_0 S_M} \bigg] \mathcal{F}(\boldsymbol{\rho}_E) \ \text{where} \ \mathcal{F}(\boldsymbol{\rho}_E) = - \frac{\mu^*\cdot \textbf{E}(\boldsymbol{\rho}_E; \beta(\omega_E))}{\sqrt{\max [\varepsilon(\boldsymbol{\rho}) |E(\boldsymbol{\rho}; \beta(\omega_E))|^2]}}
\end{align}
where $\mathcal{F}(\boldsymbol{\rho}_E)$ is a measure of the overlap of the emitter's dipole with the waveguide mode, and $S_M$ is the mode area:
\begin{align}
    S_M = \frac{\int_\Gamma \varepsilon(\boldsymbol{\rho})|\textbf{E}(\boldsymbol{\rho}; \beta(\omega_E))|^2 \textrm{d}^2\boldsymbol{\rho}}{\text{max}[\varepsilon(\boldsymbol{\rho})|\textbf{E}(\boldsymbol{\rho}; \beta(\omega_E))|^2]}
\end{align}
Similar to the cavities, the coupling between waveguides and emitters can be made stronger by reducing the mode area $S_M$ or by increasing the spatial or polarization overlap between the emitter dipole and the waveguide mode. The full hamiltonian for the waveguide-emitter system can be expressed as:
\begin{align}
    H = \omega_E \sigma^\dagger \sigma + \int_{-\infty}^{\infty}\omega a^\dagger(\omega) a(\omega) \textrm{d}\omega + \int_{-\infty}^{\infty} [g\sigma a^\dagger(\omega)  + g^* \sigma^\dagger a(\omega)] \textrm{d}\omega
\end{align}
The behaviour of this system has some marked differences from that of a cavity-emitter system. One of the most fundamental difference is that excitations in emitters coupled to waveguides exponentially decay into the waveguide mode, as opposed to periodically exchanging energy with it. In particular, if the emitter is initially in an excited state and the waveguide in vacuum state, $\ket{\Psi(0)} = \ket{e} \ket{\text{vac}}$, then at time $t$, the state of the waveguide-emitter system is given by:
\begin{align}
    \ket{\Psi(t)} = \exp[-(\textrm{i}\omega_E + \gamma /2 )t]\ket{e}\ket{\text{vac}} + \int \frac{g}{\omega - \omega_E + \textrm{i}\gamma/2} \big[\exp\{-(\textrm{i}\omega_E + \gamma / 2)t\} - \exp(-\textrm{i}\omega t)  \big] a^\dagger(\omega) \ket{g}\ket{\text{vac}}\textrm{d}\omega
\end{align}
where $\gamma = \pi |g|^2$. This process is the spontaneous emission of the emitter into the waveguide mode and $\gamma$ is the spontaneous emission decay rate --- a similar process is observed when the emitter couples to bulk media (e.g. free space), although with different decay rate. The decay rate of the emitter is a function of the electromagnetic environment in which the emitter is embedded --- this is called the Purcell effect. For a given electromagnetic environment and emitter position, the purcell enhancement factor is defined as the ratio of the spontaneous emission rate of the emitter when it is embedded inside the electromagnetic environment to the spontaneous emission rate in free space. This purcell factor is proportional to the local density of states of the electromagnetic environment \cite{oskooi2013electromagnetic} at the emitter position and along the polarization of its dipole moment.

Finally, waveguides (or more generally, loss channels) can be used to drive emitters coherently --- this corresponds to initializing the loss channel into a coherent state: $\ket{\Psi(0)} = \ket{\Psi_E} \ket{\alpha(\omega)}$ where $\ket{\alpha(\omega)}$ is a coherent state as defined in Eq.~\ref{eq:coherent_state} and $\ket{\Psi_E}$ is the state of the emitter. This is a suitable model for theoretically analyzing the response of an emitter on exciting it with a laser. A closer correspondance to the semi-classical description of light-matter interaction can be established by describing the system in terms of a state $\ket{\tilde{\Psi}(t)}$ which is related to the true state $\ket{\Psi(t)}$ a unitary transformation on $\ket{\Psi(t)}$:
\begin{align}
    \ket{\tilde{\Psi}(t)} = \exp\bigg(-\int \big[\alpha^*(\omega)a_\omega \exp(-\textrm{i}\omega t) - \alpha(\omega)a_\omega^\dagger \exp(\textrm{i}\omega t)\big] \textrm{d}\omega\bigg) \ket{\Psi(t)}
\end{align}
then the initial condition translates to $\ket{\tilde{\Psi}(0)} = \ket{\Psi_E}\ket{\text{vac}}$ equivalent hamiltonian $\tilde{H}(t)$ governing the evolution of the state $\ket{\tilde{\Psi}(t)}$ is given by (refer to Appendix C of \cite{Fischer2018scattering}):
\begin{align}
    \tilde{H}(t) = H + [\Omega^*(t) \sigma + \Omega(t) \sigma^\dagger]
\end{align}
where $\Omega(t)$ is the `driving pulse' given by:
\begin{align}
    \Omega(t) = g^* \int_{-\infty}^\infty \alpha(\omega) \exp(-\textrm{i}\omega t) \textrm{d}\omega 
\end{align}
This driving term is similar to that obtained in a treatment of light matter interaction in which the optical fields are treated classically. Addition of this driving term to the hamiltonian allows us to solve a problem in which the loss channel is initially in the vacuum state --- such a problem can be numrically analyzed within the master-equation framework \cite{Fischer2016} or the scattering matrix framework \cite{Fischer2018scattering, trivedi2018few, xu2015input}.

\section{Single photons}
\label{section:single_photon}
\subsection{Schemes for single photon generation}
\label{subsection:schemes}

\begin{figure*}[tb]
  \includegraphics[width=17.5cm]{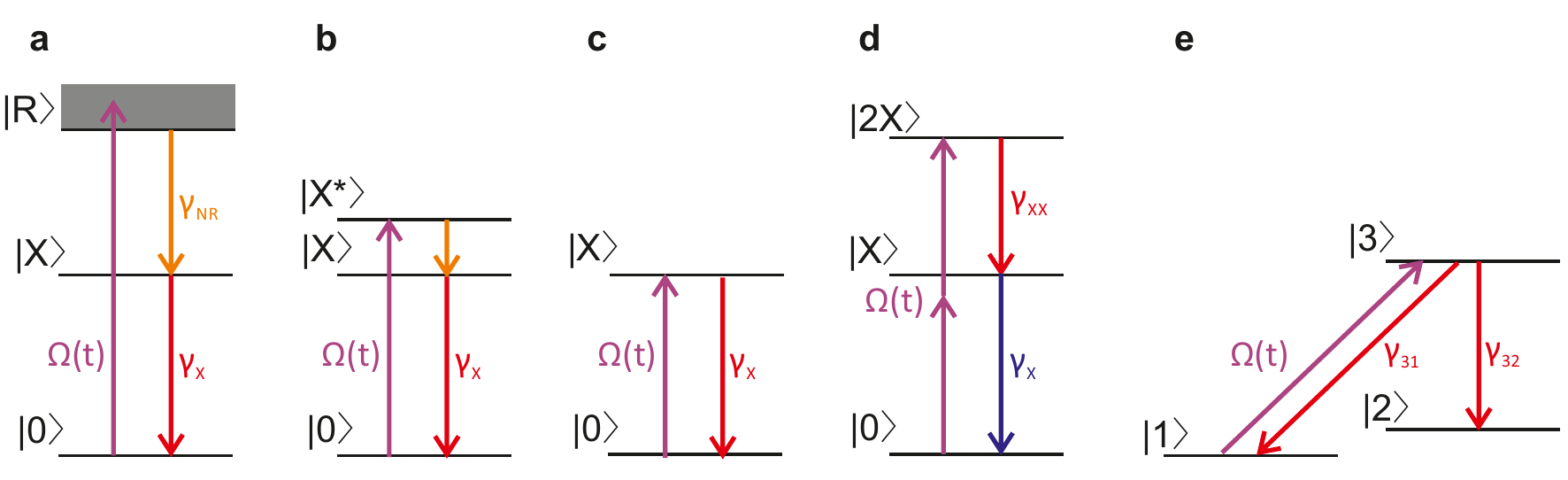}
  \caption{\textbf{Schemes for non-classical light generation: a} Non-resonant excitation above bandgap or in the wetting layer. \textbf{b} Quasi-resonant excitation via an excited state (p-shell). \textbf{c} Resonant excitation. \textbf{d} Two-photon excitation of $\ket{2X}$. \textbf{e} Lambda-scheme.}
  \label{figure:3a1}
\end{figure*}

The type of non-classical light that is most commonly generated using quantum dots are single photons. To these ends, multiple schemes with different advantages and disadvantages have been established over the years. These schemes are summarized in Fig.~\ref{figure:3a1}. They are \textbf{a}: non-resonant excitation above the bandgap, \textbf{b}: quasi-resonant excitation of an excited state (p-shell), \textbf{c}: resonant excitation, \textbf{d}: two-photon excitation of the bi-exciton, \textbf{e}: coherently excited three-level lambda system.

In the non-resonant excitation scheme (Fig.~\ref{figure:3a1}\textbf{a}) a short laser pulse creates electrons and holes either above the bandgap or in the wetting layer of the QD \cite{Michler2000, Santori2001}. These charge carriers relax into the lowest energy states of the QD via phonon-mediated processes where they recombine radiatively. If the QD is filled with multiple charge carriers, due to the Coulomb interaction each charge configuration emits at a different frequency and per excitation pulse only one photon is expected to be emitted at the energy of the neutral exciton X. This is the most simple method for generating single photons from QDs since the energy of the excitation laser does not have to be controlled precisely. Moreover, since the energy between excitation and emission is significantly different the signal can easily be separated from the excitation laser by frequency filtering. In this scheme, the purity is limited by charge carriers that stay long enough in the wetting layer / above bandgap to allow for a refilling of the QD and emission of a second photon after a first photon has been emitted \cite{Flagg2012}. Also, the degree of indistinguishability that can be achieved is limited due to the excitation timing jitter from the incoherent relaxation \cite{Fischer2016}.

To suppress re-excitation, the QD can also be excited quasi-resonantly via its p-shell (Fig.~\ref{figure:3a1}\textbf{b}). This has been shown to improve the purity \cite{Miyazawa2016}. However, the indistinguishability is still limited by the excitation timing jitter from the incoherent relaxation and although the single-photon purity is better than in scheme \textbf{a} it is still not perfect since if the relaxation and emission of a photon occur already during the presence of the excitation laser pulse the system can be re-excited and emit a second photon \cite{Fischer2016}. Depending on the ratio of the rate of the incoherent relaxation and the photon emission the limit of $g^{(2)}[0]$ will be between those of scheme \textbf{c} and \textbf{d} discussed below.

\begin{figure*}[!t]
  \includegraphics[width=13cm]{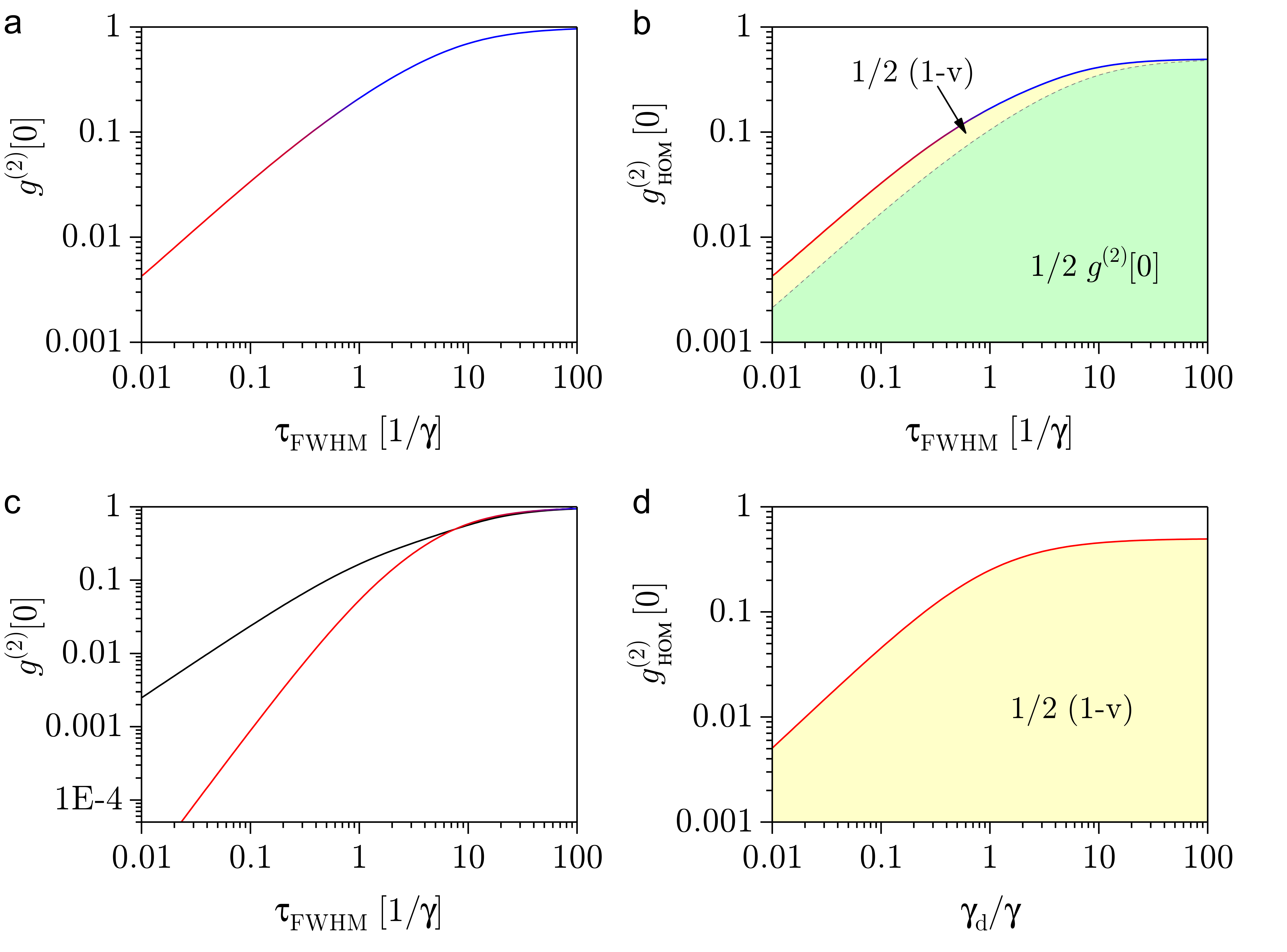}
  \caption{\textbf{Dynamics of single-photon generation. a} $g^{(2)}[0]$ as a function of the pulse length for a resonantly driven two-level system. \textbf{b} $g^{(2)}_{HOM}[0]$ as a function of the pulse length for a resonantly driven two-level system. \textbf{c} $g^{(2)}[0]$ as a function of the pulse length for a resonantly driven two-level system (black) and two-photon excitation of a three-level system (red) . \textbf{d} $g^{(2)}_{HOM}[0]$ for a lambda system as a function of the dephasing rate. (a, b and d adopted from \cite{Fischer2016}, c adopted from \cite{Hanschke2018})}
  \label{figure:3a2}
\end{figure*}

To completely eliminate excitation timing jitter resonant excitation can be performed (Fig.~\ref{figure:3a1}\textbf{c}) \cite{He2013}. Since the excitation laser and signal have the same energy here the excitation laser needs to be filtered out either by using cross-polarized detection or by using different spatial modes. This scheme has been widely used over the past years \cite{He2013, Unsleber2015, Unsleber2016, Somaschi2016, Ding2016, Wang2016, Loredo2016} and regularly demonstrated $g^{(2)}[0]<0.01$ and high indistinguishability with $v>0.97$ without Purcell enhancement of the emission rate \cite{He2013} and visibility $v>0.99$ with Purcell enhancement \cite{Somaschi2016} (see subsection \ref{subsection:resonators} below) . The finite values of the $g^{(2)}[0]$ and $1-v$ result from re-excitation \cite{Fischer2016, dada2016indistinguishable}. If a photon is already emitted during the presence of the excitation laser pulse there is a finite probability that the two-level system is re-excited and emits a second photon \cite{Fischer2017, Fischer2017-2}. The value of $g^{(2)}[0]$ for a resonantly driven two-level system is presented in Fig.~\ref{figure:3a2}\textbf{a} as a function of the pulse length (normalized to the excited state lifetime) for exciting with $\pi$-pulses of Gaussian shape. For short pulses, $g^{(2)}[0]$ increase linearly with the pulse length and saturates at 1 for long pulses. The value of $g^{(2)}_\text{HOM}[0]$ for a resonantly driven two-level system is presented in Fig.~\ref{figure:3a2}\textbf{b} as a function of the pulse length (normalized to the excited state lifetime) for exciting with $\pi$-pulses of Gaussian shape. It is decomposed into  $ g^{(2)}[0]/2$ and $(1-v)/2$ and also increases linearly for short pulses.

Recently, another scheme was used to demonstrate very pure single photon generation with values $g^{(2)}[0] \approx 10^{-5}$ \cite{Schweickert2018, Hanschke2018}. In this scheme (figure \ref{figure:3a1}d) a two-photon excitation process coherently prepares the system in the $2X$ state \cite{Brunner1994, Jayakumar2013, Ardelt2014}. Since due to the Coulomb interaction the energy of $2X$ is lower than twice the $X$ energy the energy of the excitation laser is detuned from the $2X$ as well as the $X$ transition. Single photons can then be obtained by frequency filtering either on the $X$ or $2X$ transition. Since re-excitation is only possible after the radiative cascade has returned the system to the ground state it is largely suppressed \cite{Schweickert2018, Hanschke2018}. The dependence of $g^{(2)}[0]$ on the pulse length is presented in figure \ref{figure:3a2}c for a resonantly driven two-level system (black) and two-photon excitation of a three-level system (red) for exciting with square pulses. The two-photon excitation of a three-level system shows significantly lower values than the resonantly excited two-level system. Here, we used $\gamma_{2X} = 2 \gamma_{X}$. Specifically, for short pulses $g^{(2)}[0]$ scales with $(\gamma\tau)^2$ which results for short pulses in an improvement of $g^{(2)}[0]$ as large as several orders of magnitude compared to a resonantly excited two-level system.

Another scheme that was recently demonstrated with QDs consists of a lambda system with two ground states ($\ket{1}$ and $\ket{2}$) and one excited state ($\ket{3}$) whereby the decay rate $\gamma_{32}$ is much larger than $\gamma_{31}$ (figure \ref{figure:3a1}e) \cite{Sweeney2014, Pursley2018, lee2017multi, Beguin2018}. After the system is initialized in the state $\ket{1}$ a laser pulse drives the system into state $\ket{3}$ from where it decays to state $\ket{2}$ emitting a photon. Such a system can be realized by a singly-charged QD in a Faraday geometry magnetic field where the transition 3-1 should not be optically active but in fact weakly couples to light while the transition 3-2 couples to circularly polarized light \cite{Beguin2018}. Alternatively, a singly-charged QD in a Voigt geometry magnetic field can be used. Here the energy levels consists of a double lambda system where the vertical transitions couple to vertically polarized light while the diagonal transitions couple to horizontally polarized light. Using a vertically polarized cavity mode which is slightly detuned from the QDs transition results in the scenario depicted in figure \ref{figure:3a1}e \cite{Sweeney2014, Pursley2018, lee2017multi}. In this case the single photons are Raman photons which are emitted into the cavity mode without populating the excited state of the QD. Since in this scheme the initial state and final state are different re-excitation should be entirely suppressed leading to $g^{(2)}[0]=0$ as long as the spin relaxation rate is much longer than the pulse length. Therefore, also the indistinguishability is only limited by the spin-dephasing of the ground states as well as coupling to acoustic phonons. The dependence of $g^{(2)}_{HOM}[0]$ on the dephasing rate is presented in figure \ref{figure:3a2}d. A major advantage of this scheme is that it allows to create single photons where the waveform can be arbitrarily controlled by the temporal shape of the excitation laser pulse \cite{Pursley2018, lee2017multi, Beguin2018}. The disadvantage of this scheme is that it is difficult to implement and that the rate of single-photon generation will be limited by the time that it takes to re-pump the system into state $\ket{1}$.

In addition to the inherent limitations of ideal few-level systems discussed so far QDs are subject to limitations that result from experimental constraints or the semiconductor environment. For example cross-polarized techniques that filter out the excitation laser also reduce the brightness of the signal. On the other hand, coupling to acoustic phonons in the environment damps Rabi oscillations which reduces the probability of exciting the systems and, thus also reduces the brightness. Since coupling to acoustic phonons depends on the amplitude of the driving laser, this effect is more pronounced for shorter pulses. In addition, fluctuations in the electronic environment lead to spectral diffusion of the emission frequency which further limits the obtainable indistinguishability. This effect is less pronounced in resonant excitation schemes which avoid the creating of free charge carriers. In general, a higher indistinguishability is observed for QDs embedded in diode structures which, as discussed above show less spectral diffusion due to the stabilization of the electronic environment.

\begin{figure*}[!t]
  \includegraphics[width=17.5cm]{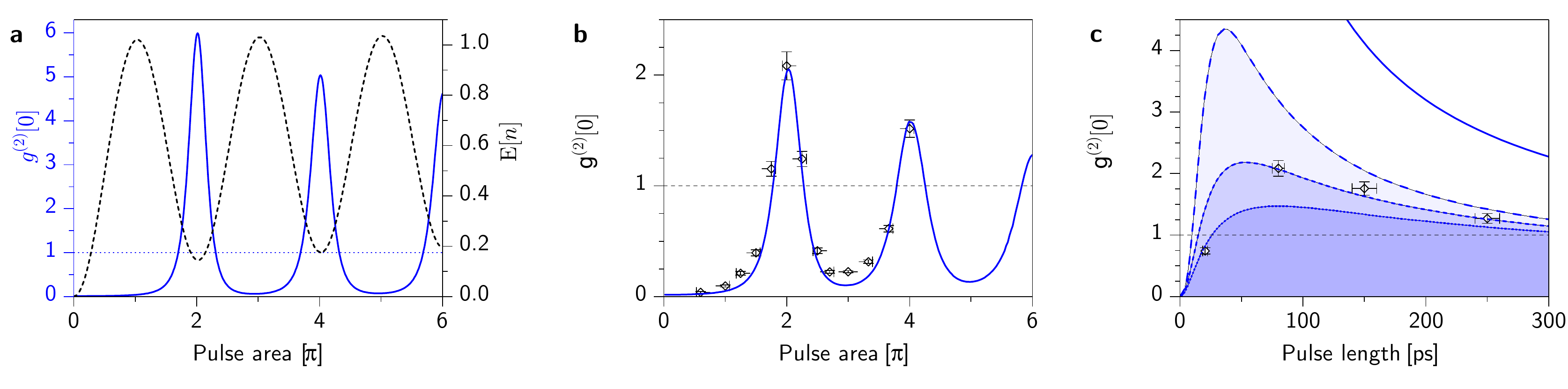}
  \caption{\textbf{Generation of two-photon pulses from a resonantly driven two-level system. a} Theoretical $g^{(2)}[0]$ (blue solid) and expected photon number (black dashed) versus pulse area for an ideal two-level system. Oscillations between anti-bunching (at odd $\pi$-pulses) and bunching (at even $\pi$-pulses) are observed. \textbf{b} Experimentally measured $g^{(2)}[0]$ versus pulse area confirming the prediction from a. Blue curve represents quantum simulations using the experimentally measured lifetime, pulse length, phonon coupling and chirp of the laser pulses. Dashed black line represents statistics of the incident laser pulse. \textbf{c} Experimental second-order coherence measurements $g^{(2)}[0]$ versus pulse length (using 2$\pi$-pulses for excitation). Solid blue line represents emission from an ideal two-level quantum system, long dashed blue line represents inclusion of dephasing, short dashed blue line represents addition of a $2.7\%$ chirp in bandwidth, and short dotted blue line represents addition of a further $2.7\%$ chirp in bandwidth. (Figure adopted from \cite{Fischer2017})}
  \label{figure:3a3}
\end{figure*}

In the discussion of the generation of single photons from a resonantly driven two-level system above we have considered only excitation pulses with a pulse area of $\pi$. However, it was recently shown that also the pulse area has a strong impact on the emitted quantum light and that a resonantly driven two-level system can even predominantly emit two-photon pulses \cite{Fischer2017}. For an ideal two-level system the expected number of emitted photons and $g^{(2)}[0]$ are presented in figure \ref{figure:3a3}a as a function of the pulse area for pulses of length $\tau=1/(10\gamma)$. Oscillations are observed in $g^{(2)}[0]$ which are out of phase with the Rabi oscillations and which can be understood as follows. For exciting with pulse areas of odd multiples of $\pi$ the emission consist mainly of single photons whereby the non-zero value of $g^{(2)}[0]$ results from the emission of a photon and re-excitation during the presence of the laser pulse as discussed above. For even multiples of pi, the emission consists mainly of the vacuum state since the Rabi oscillation returns the system to the ground state. However, if emission occurs it is most likely to occur when the system is in its excited state - exactly after an area of $\pi$ has been absorbed. Then, the remaining area of the pulse is also $\pi$ i.e. the system is re-excited and emits a second photon with very high probability which results in the observed bunching of $g^{(2)}[0]$. An experimental measurement for exciting a two-level system with a lifetime of 602ps (formed by a charged QD) with 80 ps long pulses is presented in figure \ref{figure:3a3}b and clearly confirms the theoretical prediction. Here, the blue line is a fit which takes into account the measured lifetime, measured pulse length and measured non-idealities (phonon coupling, chrip of the laser pulse). The dependence of the bunching on the pulse length  for exciting with $2\pi$ pulses is presented in figure \ref{figure:3a3}c. When taking into account the measured deviations from an ideal two-level system (different lines) a good agreement between theory and experiment is observed further confirming that a resonantly driven two-level system can preferentially emit two-photon pulses.

\subsection{Resonators (weak coupling regime)}
\label{subsection:resonators}
Resonators are commonly used to redistribute the emission such that it can be efficiently collected. Moreover, resonators with small mode volume and high quality factors (Q-factors) enhance the light-matter coupling which results in an increase of the emission rate by Purcell-enhancement and facilitates higher operational rates. Resonators with high Q-factors work only with a narrow bandwidth which is compatible with most schemes for the generation of single photons but but may be problematic for the generation of other types of non-classical light. An enhancement of the emission rate reduces the indistinguishability under non-resonant excitation since the timing jitter from the excitation process becomes more important \cite{ Kiraz2004}. On the other hand in schemes with resonant excitation increasing the emission rate is expected to increase the indistinguishability  since the homogeneous linewidth is increased relative to the inhomogeneous broadening due to spectral diffusion \cite{He2013}. However, it has has also been shown that the proximity to etched surfaces increases spectral diffusion \cite{Liu2018fab} which poses a lower limit on the resonator size. Here, surface passivation techniques may play an important role in the future \cite{Liu2018fab, Guha2017}.

Over the years, a wealth of different resonators has been developed. Prominent examples are micro-pillars, oxide aperture cavities, photonic crystal cavities, bulls-eye resonators, micro-lenses, tapered nanowires and tunable Fabry-Perot resonators. In the following, we will first discuss aspects that are relevant for all resonators before briefly describe the different resonators and highlight recent progress in the individual structures.

For all resonators, the spatial alignment between QD and resonator is important to achieve an efficient coupling between QD and resonator mode \cite{Hennessy2004}. This can be achieved either probabilistically by fabricating a large number of QD-resonator systems and then characterizing them all or deterministically by pre-characterizing QDs and then subsequently fabricating resonators around the pre-characterized QDs. Moreover, for high-Q resonators a spectral alignment of QD and resonator mode is essential \cite{Badolato2005}. Therefore, it is often useful to fine tune the detuning between QD and resonator mode and several techniques have been developed. Prominent examples are temperature tuning\cite{Yoshie2004,Reithmaier2004,Peter2005}, electric tuning of the QD emission frequency via the DC stark \cite{Strauf2007, Laucht2009} effect or ultra-fast tuning via the AC stark effect \cite{Bose2012}, deposition of monolayers \cite{Strauf2006} or condensation of inert gases \cite{Mosor2005} on the surface which changes the resonator frequency  and strain tuning, either DC \cite{Sun2013, Kremer2014} or dynamically using surface acoustic waves \cite{Fuhrmann2011}.

When resonant excitation with cross-polarized suppression of the excitation laser is combined with QDs coupled to resonators in the weak coupling regime the emission mainly consist of coherent scattering from the resonator \cite{Mueller2016}. Therefore, the structure must allow that the polarization of the coherent scattering from the resonator is not rotated, such that in can be suppressed in the detection channel,  while the polarization of the QD emission is rotated. This can be achieved, for example, in bi-modal cavities and charge-neutral QDs that have their symmetry axis different from the cavity and laser \cite{Giesz2016} or bi-modal cavities and charged QDs.

\textbf{Micropillar resonators} \cite{Pelton2002, Santori2002, Dousse2008, Heindel2010, Gazzano2013, Nowak2014, Giesz2015, Unsleber2015, Somaschi2016, Unsleber2016, Ding2016, Wang2016, Loredo2016, Schlehahn2016electric, He2017, lee2017multi} consist of a pillar with a typical diameter of a few micrometer. A $\lambda$ cavity is sandwiched between bottom and top distributive Bragg reflectors (DBR) whereby the bottom-DBR has a higher number of periods than the top-DBR to make sure that most of the emission leaves the pillar through the top. They are commonly used for efficient single photon generation since they simultaneously offer Purcell enhancement ($F_P\approx3-10$) and a high collection efficiency. The reported brightness is $0.79\pm0.08$ for non-resonant excitation \cite{Gazzano2013} and $0.37\pm0.02$ for resonant excitation\cite{Unsleber2016}, which is lower due to the polarization filtering of the emission to suppress the excitation laser and damping of the Rabi oscillations which reduces the preparation fidelity of the excited state. In order to deterministically align micropillar resonators with QDs an in-situ laser lithography technique \cite{Dousse2008} as well as a technique which is based on imaging of the QD emission \cite{He2017} have been established. Electrically contacted micropillars were demonstrated \cite{Heindel2010, Nowak2014} which further benefit from all the advantages of electrical contacting discussed in section \ref{subsection:QD} and which were used to demonstrate a very high indistinguishability with $v>0.99$ \cite{Somaschi2016}. 

\textbf{Oxide aperture cavities} \cite{Stoltz2005, Strauf2007, Rakher2009, Bonato2012, Bakker2013, Bakker2014, Bakker2015, Bakker2015OL, Snijders2016, Snijders2018PRApp, Snijders2018} are a lot like micropillar resonators with the difference that the pillar has a larger diameter of $\approx 30 \mu m$. However, the mode is confined to a diameter similar to micropillars using an oxide aperture above the QD layer resulting in comparable Purcell factors $F_P\approx2-11$. Electrically contacted oxide aperture cavities have been established in 2007 and reported values of the source brightness are as large as $0.38$ \cite{Strauf2007}. Enabled by the large diameter of the pillar a device where fibers are directly attached to the top of the pillar for efficient detection and bottom of the sample for excitation was demonstrated \cite{Snijders2018PRApp}. Thereby, the brightness of the device was  $0.05\pm0.01$ photons in the detection fiber per excitation pulse. However, this was limited by a spectral mismatch of QD and resonator and the coupling efficiency between cavity mode and fiber was measured to be $0.85\pm0.11$ and the theoretical maximums was calculated to be $0.9\pm0.076$.

\textbf{Bulls-eye resonators} \cite{Davanco2011, Ates2012, Sapienza2015, Sapienza2017} are circular DBR resonators which are fabricated by etching circular trenches into a thin membrane. The measured Purcell factor of these devices are $F_P\approx2-4$ while higher factors around $F_P\approx11-12$ were predicted for perfect spatial and spectral matching of QD and resonator mode. For this alignment, a technique based on imaging of the QD luminescence has been demonstrated  \cite{Sapienza2015, Sapienza2017, Liu2017}. The measured brightness is $0.48\pm0.05$ using an numerical aperture of $NA=0.4$ while a collection efficiency of $>0.8$ is predicted for using $NA=0.9$.

\textbf{Photonic crystal cavities} \cite{Englund2005, Faraon2007, Faraon2011, Madsen2014, Sweeney2014, Bentham2015, Pursley2018, Liu2018, Katsumi2018} consist of a thin membrane which is periodically patterned with air holes and where a deviation from the periodicity results in cavities. Due to their high quality factors and small mode volumes they are well suited for enhancing the light-matter interaction and generating non-classical light in the weak coupling regime as well as in the strong coupling regime (c.f. subsection \ref{subsection:strong-coupling}). In general, photonic crystal cavities do not have a high collection efficiency since very often a high Q-factor is achieved by reducing the cavity losses through destructive interference in the far field. However, it was recently demonstrated that for the L3 cavity, which consist of three missing air holes in a row in a hexagonal pattern, the higher-order mode M3 has a collection efficiency of $0.443\pm0.021$. Moreover, the planar geometry of photonic crystals makes them ideally suited for coupling to waveguides and on-chip photonic circuits \cite{Faraon2007, Faraon2011, Bentham2015}. To these ends, recently the generation of single photons in a cavity coupled to a waveguide was demonstrated with a very high Purcell enhancement of $F_P=43\pm2$ and high indistinguishability ($v=0.939\pm0.033$) \cite{Liu2018}.

\textbf{Microlenses} \cite{Gschrey2015, Thoma2015, Schlehahn2015, Schlehahn2016, Heindel2017, Thoma2017, Fischbach2017, Sartison2018} are micrometer sized lenses which are monolithically fabricated into the bulk substrate. The reported brightness is $0.23\pm0.03$ using a numerical aperture of $NA=0.4$ and collection efficiencies $>0.8$ are expected using more sophisticated lens designs \cite{Gschrey2015}. Moreover, a collection efficiency of $0.4\pm0.04$ has been demonstrated using the combination of a monolithic microlens combined with a 3D printed micro-objective \cite{Fischbach2017}. Since microlenses are not high-Q resonators which enhance the emission rate they operate broadband. Spatial alignment between QD and lens has been realized using an in-situ cathode luminescence imaging and electron beam lithography technique \cite{Gschrey2013, Gschrey2015} as well as in-situ optical lithography\cite{Sartison2018}.

\textbf{Tapered nanowires} \cite{Claudon2010, Reimer2012, Munsch2013, Kremer2014, Reimer2016, Chen2016} are nanowires which are tapered to adiabatically expand the guided mode towards the top of the nanostructure which results in a high out-coupling and collection efficiency. Similar to microlenses, the absence of a high-Q resonator facilitates broadband operation. The nanowire can either be fabricated top-down \cite{Claudon2010} or bottom-up during the growth of nanowire QDs \cite{Reimer2012} whereby the bottom-up process automatically ensures spatial alignment between QD and resonator. Interestingly, the taper can either decrease or increase the width towards the top and the reported brightness is $0.75\pm0.1$ \cite{Munsch2013}.

\textbf{Tunable Fabry-Perot resonators}
Tunable Fabry-Perot resonators \cite{Muller2009, Miguel-Sanchez2013, Greuter2015, Najer2017, Najer2018, Herzog2018} consist of two parts: the sample which consist of a planar back DBR as well as a part of the cavity containing the QDs and the curved top DBR which is either fabricated into the tip of a fiber \cite{Muller2009, Miguel-Sanchez2013, Herzog2018} or a glass template \cite{Greuter2015, Najer2017, Najer2018}. Therefore, they allow to easily tune the resonator frequency by changing the distance between bottom part and top part. Moreover, by moving the bottom part and top part relative to each other in lateral direction it is possible to spatially align any QD in the bottom part with the resonator mode. Due to the high Q-factor and small mode volume, this type of resonator was recently used to demonstrate strong coupling very far in the strong coupling regime \cite{Najer2018}. While efficient single photon generation has not been demonstrated so far, calculations indicate that for lower Q-factor resonators extraction efficiencies of $>0.9$ are within reach\cite{Najer2018}.

\subsection{Resonators (strong coupling regime)}
\label{subsection:strong-coupling}

\begin{figure}[!t]
  \includegraphics[width=8.6cm]{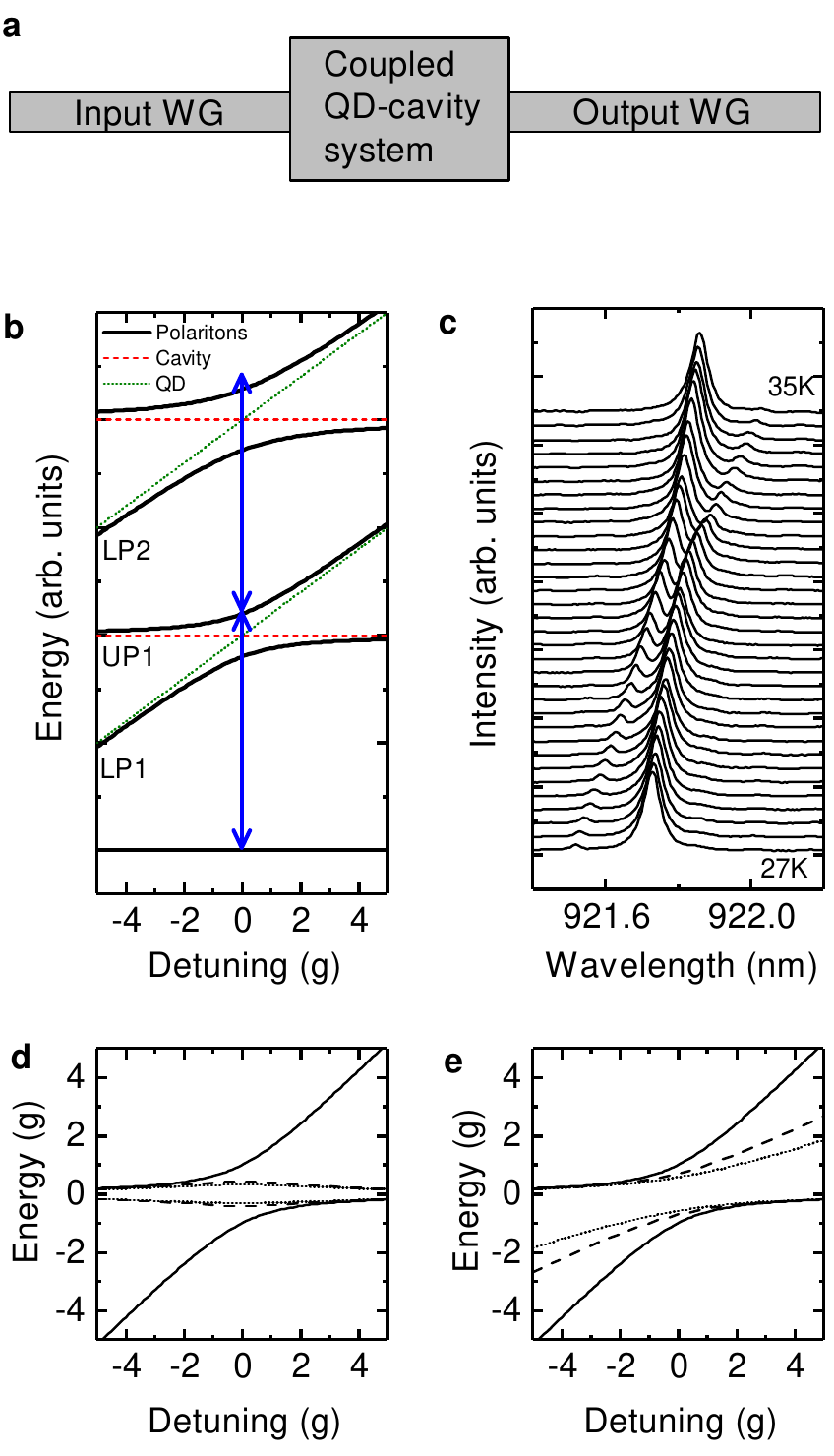}
  \caption{\textbf{Generation of single photons from strongly coupled systems. a} Illustration of on-chip architecture. \textbf{b} Schematic illustration of the first two-rungs of the Jaynes-Cummings ladder and visualization of photon blockade. \textbf{c} Anticrossing of the lowest rung measured in cross-polarized reflectivity. \textbf{d} Energy of transitions from the ground state to the first rung (solid lines) and subsequent climbs of higher rungs (dashed and dotted lines). \textbf{e} Energies of exciting the $n$-th rung of the JC-ladder in an $n$-photon process for $n=1$ (solid lines), $n=2$ (dashed lines) and $n=3$ (dotted lines). In d and e the energies are given relative to the cavity energy. (b, d and e adopted from \cite{Mueller2015}, c adopted from \cite{Mueller2015-2})}
  \label{figure:3c1}
\end{figure}

As discussed above, for off-chip applications resonant excitation can be combined with a QD weakly coupled to a resonator if the structure allows that the polarization of the coherent scattering from the resonator is not rotated, such that in can be suppressed in the detection channel,  while the polarization of the QD emission is rotated. However, for on-chip applications, where a QD-resonator system is coupled to an input waveguide and an output waveguide (figure \ref{figure:3c1}a) the transmission consists mainly of coherently scattered laser and is far away from single photons. While experiments have been performed with this on-chip transmission geometry, the same physics can be investigated using an off-chip transmission geometry which consists of cross-polarized resonant excitation and detection whereby the linearly polarized resonator mode is aligned diagonal to the setup polarization axis.

In transmission geometry, single photons can be generated using a strongly coupled QD-cavity system, i.e. when the coupling strength $g$ exceeds the QD emission rate $\gamma$ and cavity loss rate $\kappa$. When QD and cavity are tuned into resonance the eigenstates of the system are polaritons with energies $E_n^{\pm}=n\hbar\omega\pm 2\sqrt{n}\hbar g$ where $n$ is the number of phonons involved and which is known as Jaynes-Cummings ladder (JC-ladder). The energy structure as a function of the QD-cavity detuning $\Delta$ is visualized in figure \ref{figure:3c1}b for $n\leq2$. At resonance, the eigenstates of each rung $n$ are the polaritons which are split by $\sqrt{n}2g$ and with increasing detuning they evolve towards the bare QD and cavity states. The characteristic anticrossing of the lowest rung can be measured using photoluminescence \cite{Reithmaier2004, Yoshie2004, Peter2005} or cross-polarized reflectivity \cite{Englund2007} as presented in figure \ref{figure:3c1}c \cite{Mueller2015-2}. In this system single photons can be generated through the anharmonicity of the JC-ladder. A laser tuned into resonance with the first rung of the ladder (blue arrow figure \ref{figure:3c1}c) is not in resonance with subsequent climbs up the ladder. Therefore, the admission of photon to the system reduces the probability for a second photon to be admitted which is known as photon blockade\cite{Faraon2008, Reinhard2011}. However, due to the highly dissipative character of nanophotonic systems the observed antibunching for QD and cavity in resonance is moderate \cite{Faraon2008, Reinhard2011}. Although a laser in resonance with the first rung is not in exact resonance with the transmission from the first rung to the second rung it still has significant overlap with the transition due to the linewidth of the states. In order to increase the anharmonicity and reduce the value of $g^{(2)}[0]$ detuning of QD and cavity can be used \cite{Mueller2015}. This is visualized in figure \ref{figure:3c1}d  which presents the detuning dependent energy differences between the rungs for transitions leading to $n=1$, $n=2$ $n=3$ as solid, dashed and dotted lines respectively. With detuning the transition energies to the first rung and subsequent climbs up the JC-ladder increases for exciting the more QD-like polariton. This is also the case for the energies of exciting the $n$-th rung of the JC-ladder in an $n$-photon process (figure \ref{figure:3c1}e). In addition, the pulse length of the excitation laser can be optimized \cite{Mueller2015-2}: longer excitation pulses increase the chance of re-excitation during the presence of the pulse while shorter pulses have a broader linewidth and thus increase the overlap with the different transitions. Exploiting detuning of QD and cavity and optimizing the pulse length allowed to observe antibunching with $g^{(2)}[0]=0.34\pm0.07$. Importantly, the limiting factor is coherent scattering of the laser from the detuned cavity \cite{Mueller2016}.

\begin{figure}[!t]
  \includegraphics[width=8.6cm]{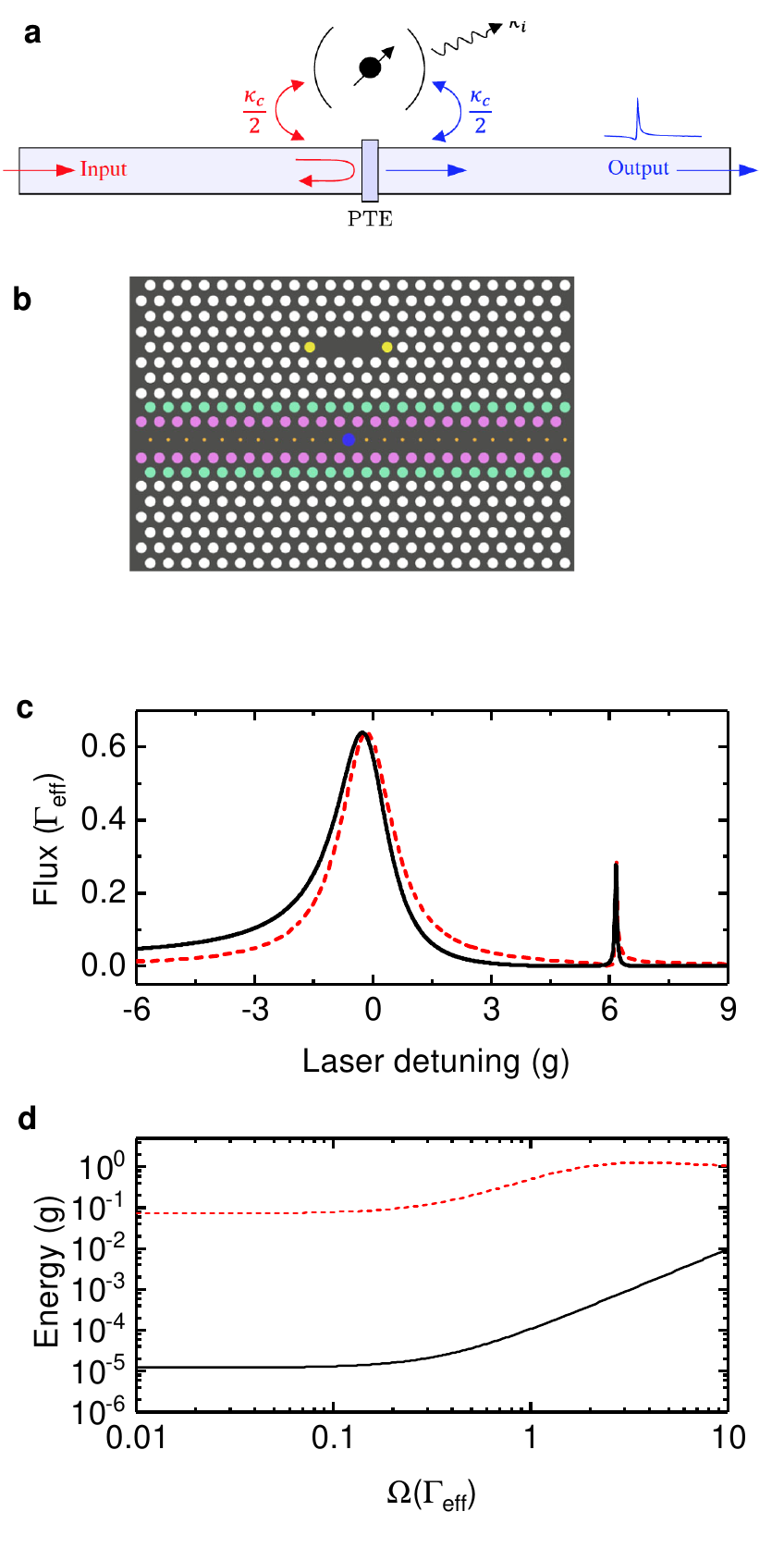}
  \caption{\textbf{Interferometric cancellation of coherent scattering from the cavity. a} Schematic illustration of a cavity coupled to an input and an output waveguide and where a direct transmission from input to output waveguide can be controlled with a PTE. \textbf{b} Suggested structure based on a photonic crystal platform. Colored circles indicate holes where the position or size is changed relative to the crystal. \textbf{c} Transmission of the system for a QD-cavity detuning of $\Delta=6g$ with the PTE fully blocking (red dashed line) and optimized for interferometric cancellation (black solid line). \textbf{d} $g^{(2)}[0]$ as a function of the driving strength for the laser in resonance with the more QD-like polariton with the PTE fully blocking (red dashed line) and optimized for interferometric cancellation (black solid line). Figure adopted from \cite{Fischer2016-3}}
  \label{figure:3c2}
\end{figure}

For a photonic crystal cavity system in off-chip geometry this coherent scattering of the laser from the detuned cavity can be interferometrically cancelled using a self-homodyne effect which is intrinsic to photonic crystal cavities \cite{Fischer2016-2}. In this effect the coherent scattering from the red detuned cavity interferes with scattering from far blue detuned higher-order cavity modes and above-the-lightline modes which occurs with opposite phase. Consequently, by balancing the relative strength of these two components the coherent scattering from the cavity can be entirely suppressed. This allows to isolate the non-classical light and has enabled the observation of single-photon generation with $g^{(2)}[0]=0.05\pm0.04$, $v=0.96\pm0.05$ and an emission rate of $1/55 ps^{-1}$ \cite{Mueller2016}. To realize this interferometric cancellation in an on-chip transmission geometry a sample design based on an input and output waveguide which are coupled not only to the cavity but also to a partially transmitting element (PTE) can be used (figure \ref{figure:3c2}a) \cite{Fischer2016-3}. The simulated transmission of such a proposed structure (figure \ref{figure:3c2}b) is presented in figure \ref{figure:3c2}c for $\Delta=6g$ which shows the transmission with the PTE fully blocking (JC-system) and tuned to optimal interferometric cancellation as dashed and solid lines respectively \cite{Fischer2016-3}. Here a fabricable Q-factor of the cavity of $\sim 51.000$ in the absence of the waveguides and a easily achievable coupling strength of $g=10 \times 2\pi$ have been used \cite{Yoshie2004, Englund2007, Faraon2008, Ota2018}. The power dependent $g^{(2)}[0]$ is presented in figure \ref{figure:3c2}d for the laser in resonance with the more QD-like polariton with the PTE fully blocking and tuned to optimal interferometric cancellation as dashed and solid lines respectively. While without interferometric cancellation (dashed line) the purity of single-photon generation is moderate ($g^{(2)}[0] \approx 0.1$) with optimal interferometric cancellation (solid line) high-quality single photons with $g^{(2)}[0]<10^{-4}$ are observed for effective driving strengths $<1$ (where 1 corresponds to a pulse area of $\pi$). We would like to note here, that this interferometric cancellation of coherent scattering from a detuned cavity is not exclusive to strongly-coupled systems but similarly works for weakly-coupled systems \cite{fischer2018pulsed}.

\subsection{Waveguides}
\label{subsection:waveguides}
For the efficient generation of single photons in waveguides the QDs can also be directly integrated into waveguides without cavities \cite{Lund-Hansen2008, Thyrrestrup2010, Schwagmann2011, Laucht2012, Fattah2013, Arcari2014, Makhonin2014, Soellner2015, Javadi2015, Daveau2017, Kirsanske2017, Javadi2018, Zhou2018, Thyrrestrup2018, Hallett2018}. Here, photonic crystal W1 waveguides, which consist of one row of missing holes, have been used \cite{Lund-Hansen2008, Thyrrestrup2010, Schwagmann2011, Laucht2012, Fattah2013, Arcari2014, Javadi2015, Daveau2017, Hallett2018} as well as photonic crystal glide-plane waveguides, which allow for a chiral light-matter coupling, \cite{Soellner2015} and nanobeam waveguides \cite{Makhonin2014, Kirsanske2017, Javadi2018, Zhou2018, Thyrrestrup2018}. While QDs in waveguides do not experience Purcell enhancement, their efficient emission into the waveguide mode results in a high source brightness in the waveguide. This is quantified by the $\beta$-factor which is the ratio of emission into the desired mode compared to all emission and for QDs in photonic crystal waveguides, $\beta$-factors as high as $\beta=98.43\pm0.04$ have been observed \cite{Arcari2014}. The single photons can then be used on-chip, e.g. for quantum networks \cite{Mahmoodian2016} or efficiently coupled to single mode fibers\cite{Daveau2017, Zhou2018}. In addition to generating single photons using the schemes discussed above, also the transmission through a coupled QD - single mode waveguide system can generate single photons through nonlinear quantum optics \cite{Javadi2015, Thyrrestrup2018}.

\section{Entangled photon pairs}
\label{section:entangled}
\begin{figure}[!t]
  \includegraphics[width=8.6cm]{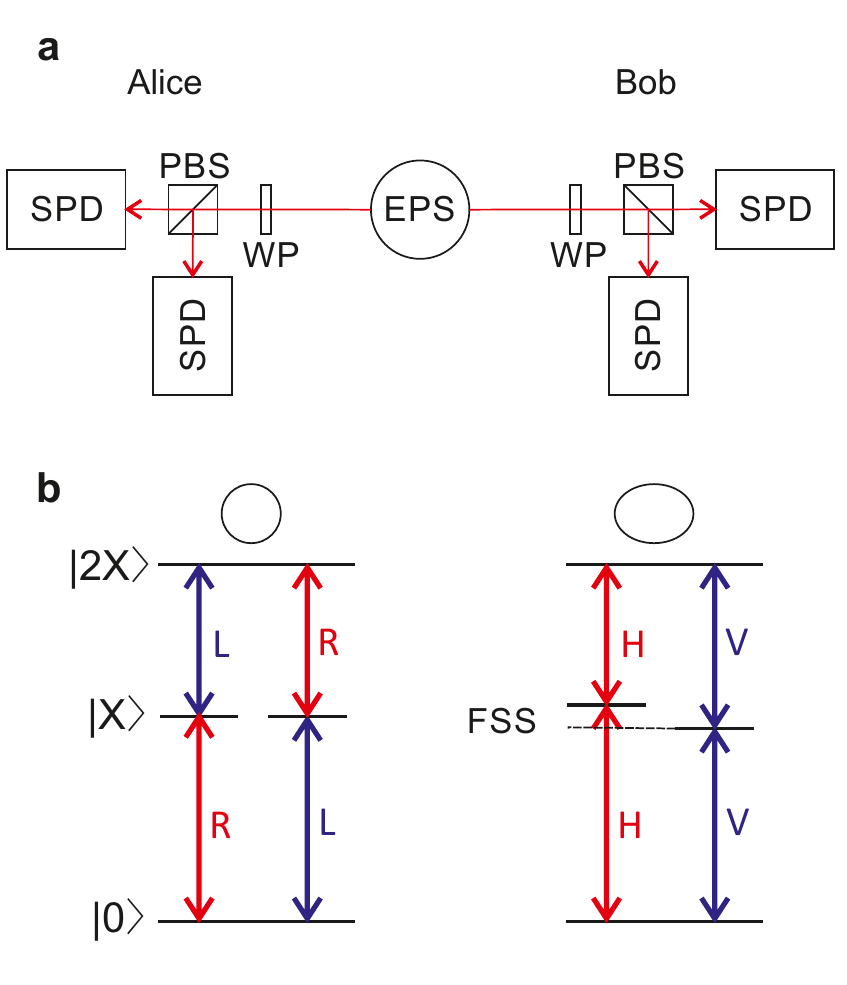}
  \caption{\textbf{Entangled photon pairs. a} Schematic illustration of a setup for using entangled photons for quantum communication and for quantifying the entanglement. EPS: Entangled photon source, SPD: Single photon detector, PBS: Polarizing beamsplitter, WP: waveplates. \textbf{b} Schematic illustration of the lowest energy levels of an uncharged QD with (left) and without (right) $D_{2D}$ symmetry.}
  \label{figure:41}
\end{figure}

Entangled photon pairs are essential for several quantum communication protocols such as the E91 protocol \cite{Ekert1991} in which an entangled photon pair source (EPS) is used for quantum key distribution (Fig.~\ref{figure:41}\textbf{a}). The EPS emits a pair of entangled photons in one of the four Bell states:
\begin{align*} 
\ket{\Phi^+} &= \frac{1}{\sqrt{2}}(\ket{H}_A\ket{H}_B+\ket{V}_A\ket{V}_B)   \\ 
\ket{\Phi^-} &= \frac{1}{\sqrt{2}}(\ket{H}_A\ket{H}_B+\ket{V}_A\ket{V}_B)   \\
\ket{\Psi^+} &= \frac{1}{\sqrt{2}}(\ket{H}_A\ket{V}_B+\ket{H}_A\ket{V}_B)   \\ 
\ket{\Psi^-} &= \frac{1}{\sqrt{2}}(\ket{H}_A\ket{V}_B-\ket{H}_A\ket{V}_B)   \\
\end{align*}
where $H$ and $V$ denote the polarization of the photons and the indices the spatial position Alice (A) or Bob (B). When Alice and Bob measure in the same basis the results are perfectly correlated independent of the choice of basis which can be used to generate a secret key and eavesdropping can be detected by tests of Bell's theorem.

For many applications it is important that the entanglement is of high quality. As was done for a mixed single-photon source, a density matrix can be factorized into a sum of density matrices corresponding to pure states:
\begin{equation}
\hat{\rho}=\sum_i \sigma_i \ket{\Psi_i}\bra{\Psi_i}
\end{equation}
where $\sigma_i$ denotes the portion of an ensemble to be in the state $\ket{\Psi_i}$. For example, the state $\ket{\Psi^+}$ has the density matrix
$$
\hat{\rho}_0 = 
\begin{pmatrix}
1&0&0&1\\
0&0&0&0\\
0&0&0&0\\
1&0&0&1\\
\end{pmatrix}
$$
To experimentally measure the density matrix $\hat{\rho}_\text{exp}$, 16 correlation measurements using different bases have to be performed with a setup as schematically illustrated in Fig.~\ref{figure:41}a. The degree of entanglement can then be quantified through several parameters which can be calculated from the deviation of $\hat{\rho}_\text{exp}$ from $\hat{\rho}_0$. For QDs, the most common measure is the fidelity that can be written as $F=\text{Tr} [\hat{\rho}_\text{exp}\cdot \rho_0]$ (which is the inner product of two hermitian operators associated with the Frobenius norm). Importantly, this quantity can be measured using less than the 16 correlation measurements mentioned above.

In the year 2000 it was proposed that QDs can be used as sources of entangled photon pairs \cite{Benson2000} which was for the first time experimentally demonstrated  in 2006 \cite{Akopian2006} and over recent years their potential as high-quality sources has been established \cite{Hafenbrak2007, Mueller2009, Dousse2010, Versteegh2014, Mueller2014, Trotta2014, Huber2014, Jayakumar2014, Trotta2015, Zhang2015, Chen2016entangled, Trotta2016, Huber2016, Keil2017, Olbrich2017, Huber2017, Jons2017, Zhang2017, Huber2018, Mueller2018, Basset2018, Primueller2018}. An important milestone for the generation of entangled photon pairs was the development of two-photon excitation of $\ket{2X}$ either in a resonant two-photon process \cite{Brunner1994, Jayakumar2013} or via a detuned phonon-mediated process \cite{Ardelt2014} which allows for an on-demand generation with high efficiency \cite{Mueller2014}. The energy level diagram of the lowest energy states of uncharged QDs was already discussed in section \ref{subsection:QD} and is reproduced in figure \ref{figure:41}b. When preparing a QD with $D_{2D}$ symmetry in the $\ket{2X}$, the radiative decay of the state $\ket{2X}$ will produce a photon which has its polarization entangled with the spin of the state $\ket{X}$. Subsequently, the radiative decay of $\ket{X}$ will transfer the entanglement to the second photon which results in an entangled photon pair which can be written as:
\begin{equation}
\ket{\Psi} = \frac{1}{\sqrt{2}}(\ket{R}_{2X}\ket{L}_X+\ket{L}_{2X}\ket{R}_X)
\end{equation}
Since the energies of the two-photons are different, they can be easily separated. However, any realistic QD will not have $D_{2D}$ symmetry but rather some degree of anisotropy which, as discussed in section \ref{subsection:QD}, leads to two non-degenerate $\ket{X}$ states which are split by a finestructure splitting of energy $FSS$ and couple to linearly polarized light. Therefore, the entangled wavefunction will take the form
\begin{equation}
\ket{\Psi(t)} = \frac{1}{\sqrt{2}}(\ket{H}_{2X}\ket{H}_X + e^{i \frac{FSS}{\hbar}t}\ket{V}_{2X}\ket{V}_X)
\end{equation}
and the entanglement precesses in time. This means that in a pulse-wise integrated form, which is desired for applications, the degree of entanglement degrades. Moreover, single-photon detectors have a limited timing resolution which can make it difficult to resolve the precession.

Consequently, in order to obtain a high degree of entanglement the finestructure splitting needs to be entirely removed. Removing or reducing the finestructure splitting can be done using the optical Stark effect \cite{Mueller2009}, a combination of applying strain in the plane of the QDs and vertical electric fields \cite{Trotta2012, Trotta2012am} as well as by microstructures which can control the amount of strain and direction independently \cite{Trotta2015, Chen2016entangled, Trotta2016}. Alternatively, QDs which are inherently highly symmetric such as nanowire QDs \cite{Huber2014, Versteegh2014, Jons2017} or GaAs/AlGaAs QDs grown by droplet epitaxy can be used \cite{Huo2013, Keil2017, Huber2017, Huber2018, Basset2018}. Using the latter material system entanglement fidelities of up to $0.978(5)$ have been observed \cite{Huber2018}. These QDs also have the additional benefit that their emission wavelengths are suitable for interfacing with atomic vapor quantum memories \cite{Keil2017, Huber2017}. Recently, in addition to polarization entanglement and time-bin entanglement  hyperentanglement, which is the combination of both, has been demonstrated \cite{Primueller2018}. Enhancing the extraction efficiency or emission rate is more difficult for entangled photon pairs compared to single photons. Due to the different energies of the $2X$ and $X$ transition, either broadband resonators can be used or resonators that support two modes which are tuned in resonance with the 2X and X transitions respectively \cite{Dousse2010}. Moreover, for polarization entangled photon pairs the modes need to be bi-modal to support both polarizations which is easier to achieve for low-Q resonators. Remarkably, an extraction efficiencies of $0.65\pm0.04$ was recently reported for dielectric antennas \cite{Chen2018}. 

\section{Photonic graph states}
\label{section:graph}
In quantum information processing and quantum communication, graph states are highly entangled states that can serve as a resource for many information processing tasks. A graph state $\ket{\Psi_G}$ associated with a graph $G = (V, E)$ can be constructed by associating a qubit with every vertex in $V$ and initializing it in the state $\ket{+} = (\ket{0} + \ket{1})/\sqrt{2}$ followed by an application of a controlled-Z gate between qubits whose corresponding vertices have an edge in $E$ \cite{raussendorf2003measurement}:
\begin{align}
    \ket{\Psi_G} = \prod_{(v_1, v_2) \in E} \textrm{CZ}^{(v_1, v_2)} \ket{+}^{\otimes V}
\end{align}
An alternative and equivalent definition of a graph state is as follows --- for every vertex $v \in V$, define an operator $\hat{K}^{(v)}$:
\begin{align}
    \hat{K}^{(v)} = \hat{\sigma}^{(v)}_x \prod_{u \in N_v} \hat{\sigma}_z^{(u)}
\end{align}
where $N_v$ is the neighbourhood of $v$ (i.e. $N_v = \{u |\ \exists (u, v) \in E \}$). The graph state $\ket{\Psi_G}$ is then defined as a common eigenvector of $\hat{K}^{(v)} \ \forall \ v \in V$ with unity eigenvalue:
\begin{align}
    \hat{K}^{(v)} \ket{\Psi_G} = \ket{\Psi_G} \ \forall \ v \in V
\end{align}
Photonic graph states would refer to such graph states realized with photonic qubits as the physical platform. Most proposals of photonic graph states are based on pulsed polarization qubits --- two orthogonal polarizations of light (e.g. horizontal and vertical, right and left circularly polarized) are treated as the $\ket{0}$ and $\ket{1}$ states, with different pulses being treated as different qubits.

Graph states have been shown to be an essential resource for measurement based quantum computation \cite{raussendorf2003measurement, nielsen2004optical} --- in this model of quantum computation, unlike its gate-based counterpart, all the qubits are initialized into a cluster state followed by a sequence of single qubit measurements. A specific quantum algorithm can be obtained by suitably designing the initial graph state and a sequence of such measurements. Since all the entanglement needed in the quantum computation is prepared in the initial state, this approach to quantum computation requires a smaller number of auxillary qubits as compared to gate-based quantum computation. Additionally, the use of cluster states for quantum computation obviates the need for design of two qubit gates, which are especially hard to realize for photonic qubits due to the non-interacting nature of photons.

Apart from quantum computation, graph states are becoming increasingly important in quantum communication. Quantum communication, which refers to the process of transmitting quantum information across a communication channel (e.g. an optical fiber), relies on entanglement between the states of the transmitter and receiver --- most quantum communication protocols require a maximally entangled state (e.g. Bell states) to be shared between the two ends of the communication channel \cite{nielsen2002quantum}. Generating this entanglement over long distances is challenging, and a practical quantum communication system needs refreshment of this entanglement at intermediate nodes referred to as quantum repeaters \cite{briegel1998quantum}. The conventional quantum repeater architecture has atomic quantum memories entangled with propagating single-photon wave packets at these intermediate nodes, and a long-distance entanglement these atomic quantum memories is established by performing a Bell measurement on pairs of the propagating single photons. Such architectures typically require a very large coherence time for the atomic memories, making them infeasible with current technology. As an alternative, all-photonic repeater graph states (RSG) have been proposed \cite{azuma2015all}. This RSG contains $2m$ photonic qubits out of which $m$ qubits (referred to as the $1^\text{st}$ leaf qubits) form a completely connected graph state, with the remaining $m$ qubits being (referred to as the $2^\text{nd}$ leaf qubits) each connected to one of the $1^\text{st}$ leaf qubit. Such an all-photonic state can be used as a building block of a quantum repeater, with the long distance entanglement being created by Bell measurements between $2^\text{nd}$ leaf qubits of neighbouring graph states.

Generation of photonic graph states would involve creating entanglements between photonic qubits --- this has traditionally been a challenging tasks since most optical systems are inherently linear. Nonlinearity in optical system can either be introduced through material nonlinearity (e.g. optical nonlinearities such as kerr or $\chi^{(2)}$ nonlinearity, or by coupling the optical field to emitters such as quantum dots), or through photo-detection. Photo-detection based graph states are often generated by preparing photon pairs in an entangled (Bell) state via parametric down conversion, followed by a joint detection on photons from different pairs. This photo-detection process fuses the entangled pairs together, and different graph states can be designed by controlling the sequence of measurements performed. However, due to the involved photo-detection process, this scheme of generating graph states is inherently probabilistic and suffers from scalability issues for larger graph states.

\begin{figure}
    \centering
    \includegraphics[scale=0.7]{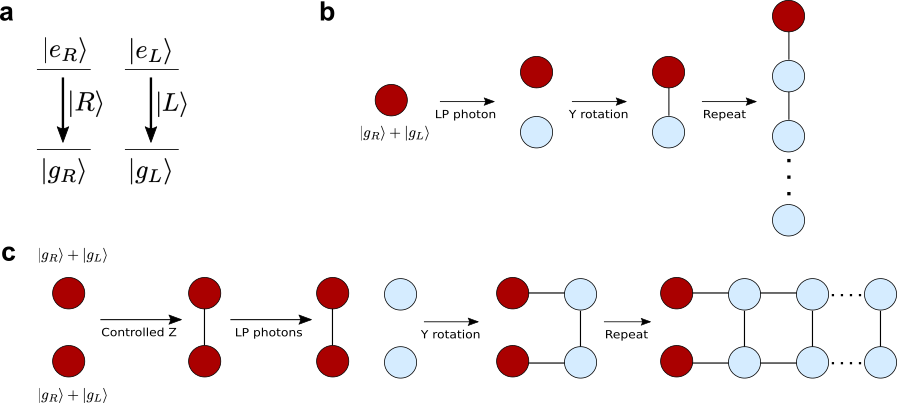}
    \caption{\textbf{Graph state generation.} \textbf{a} Level structure of an emitter that can be used to generate cluster state --- this level structure can be achieved using a singly-charged quantum dot as described in section \ref{subsection:QD}. \textbf{b} Generation of a linear graph state using a single emitter. \textbf{c} Generation of a two-dimensional ladder graph state using two coupled emitters. In both \textbf{b} and \textbf{c}, red circles denote emitter qubits (with $\ket{g_L} \equiv \ket{0}$ and $\ket{g_R} \equiv \ket{1}$ and blue circles denote photonic qubits (with $\ket{R} \equiv \ket{0}$ and $\ket{L} \equiv \ket{1}$).}
    \label{fig:cluster_state}
\end{figure}

Employing emitters to create entanglement between photonic qubits allows us to create a deterministic scheme for generating a given cluster state. The rich level structure of quantum dots and the associated selection rules makes them a potential candidate for entangling photonic qubits. Several proposals for realizing graph states that requires a level structure shown in figure \ref{fig:cluster_state}  have been made in the past decade --- this level structure can be obtained in a singly-charged quantum dot (section \ref{subsection:QD}). This level structure has two ground states $\ket{g_R}$ and $\ket{g_L}$ that coupled via optical transitions to two excited states $\ket{e_R}$ and $\ket{e_L}$ respectively --- the transition from $\ket{g_R}$ to $\ket{e_R}$ couples to a right circularly polarized optical mode and that from $\ket{g_L}$ to $\ket{e_L}$ couples to a left circularly polarized optical mode. Generation of a linear graph state is easily accomplished by applying a sequence of optical excitations and rotations between the two ground states \cite{lindner2009proposal} --- this is shown in figure \ref{fig:cluster_state}\textbf{b} and described below:
\begin{enumerate}
\item Initialize the emitter to a (unnormalized) state $\ket{g_R} + \ket{g_L}$ and excite it with a linear polarized single photon. Since a linear polarized photon will be an equal superposition of left and right circularly polarized photon, the final entangled state of the photon-emitter system would be $\ket{R}\ket{g_R} + \ket{L}\ket{g_L}$.
\item Apply a $\pi / 2$ rotation on the qubit formed by the ground states ($\ket{1} \equiv \ket{g_R}$ and $\ket{0} = \ket{g_L}$) around the $y$ axis to obtain the state $(\ket{g_L} + \ket{g_R})\ket{R}_1 + (-\ket{g_R} +\ket{g_L})\ket{L}_1$ --- this rotation can be easily performed by applying a suitable oriented magnetic field to the quantum dot. At this point, the emitter and the emitted photon already form a linear graph state with two vertices (assuming that $\ket{g_R}$ and $\ket{R}_1$ are interpreted as $\ket{1}$ and $\ket{g_L}$ and $\ket{L}_1$ are interpreted as $\ket{0}$).

\item  Scatter another linearly polarized photon from the emitter, followed by another $y$ rotation of the ground state to obtain the linear graph state with 3 vertices. Repeat this step to increase the length of the linear graph state.
\end{enumerate}
Note that the graph state created so far has 1 qubit formed by the ground states of the emitter and the remaining qubits formed by emitted the left and right polarized photon. A measurement on the most recently generated photon on the computational basis ($\ket{R}$ and $\ket{L}$) would project the emitter into one of the ground states and the remaining photons into a (smaller) linear graph state.

This proposal can be adapted to generate two-dimensional cluster states by employing coupled emitters as delineated in \cite{economou2010optically}. Figure \ref{fig:cluster_state}\textbf{c} diagramatically shows the generation of such a cluster state. Both emitters are initialized to an equal superposition of the two ground states ($(\ket{g_R}_1 + \ket{g_L}_1) \otimes (\ket{g_R}_2 + \ket{g_L}_2)$), followed by an application of the controlled-Z gate between the two emitters. This effectively creates a two qubit graph state between the two emitters ($\ket{g_R}_1 \ket{g_R}_2 - \ket{g_R}_1 \ket{g_L}_2 + \ket{g_L}_1 \ket{g_R}_2 + \ket{g_L}_1\ket{g_L}_2$). It was proposed that the CZ gate can be realized by optically coupling the trion states of the quantum dots, which lie above $\ket{e_R}$ and $\ket{e_L}$ in the energy spectrum \cite{stinaff2006optical}. Next, we scatter a linearly polarized photon off each emitter. After the completion of the scattering process, the emitters and photons would be in an entangled state $\ket{g_R}_1 \ket{R}_1 \ket{g_R}_2 \ket{R}_2 - \ket{g_R}_1 \ket{R}_1 \ket{g_L}_2 \ket{L}_2+ \ket{g_L}_1 \ket{L}_1 \ket{g_R}_2 \ket{R}_2 + \ket{g_L}_1 \ket{L}_1 \ket{g_L}_2 \ket{L}_2$. Note that this is still a two qubit graph state with $\ket{1} \equiv \ket{g_R}\ket{R}$ and $\ket{0} \equiv \ket{g_L}\ket{L}$. Application of a $\pi / 2$ rotation along $y$-axis on the qubit formed by $\ket{g_R}$ and $\ket{g_L}$ for both the emitters results in a square graph state between the four qubits (two emitter qubits and two photonic qubits). Repeating this step increases the length of the graph state, resulting a ladder graph state. A similar strategy of applying constrolled-Z gates in between two emitters followed by scattering of single photons together with the local complementation \cite{van2004graphical} property of graph states can be used to deterministically generate more complicated graph states \cite{buterakos2017deterministic, russo2018deterministic}. Recently, for the first time the generation of a 1D graph state using semiconductor QDs has been demonstrated \cite{Schwartz2016}.

\section{Conclusion}
In summary, this review discussed the generation of non-classical light using semiconductor quantum dots. We discussed the underlying fundamentals, including a thorough theoretical description of quantum light and light-matter interaction and summarized recent progress in the generation of single photons, entangled photon pairs and cluster states. This field has made an enormous progress of the past few years and high-quality sources of non-classical light based on semiconductor quantum dots have been demonstrated using many different approaches. Therefore, we expect that these sources will play a pivotal role in fundamental quantum physics research as well as quantum technology applications in the near future.

\section{Acknowledgement}
We gratefully acknowledge financial support from the German Federal Ministry of Education and Research via the funding program Photonics Research Germany (contract number 13N14846), the European Union’s Horizon 2020 research and innovation programme under grant agreement No. 820423 (S2QUIP), the Bavarian Academy of Sciences and Humanities and by the National Science Foundation grant ECCS 1839056. RT acknowledges funding from Kailath Stanford Graduate Fellowship.

\section{Bibliography}

\bibliography{BibliographyX}

\appendix
\section{Quantization of electromagnetic fields}\label{sec:em_quant}
The starting point of electromagnetic field quantization is the electromagnetic hamiltonian, which can be expressed in terms of the field $\textbf{E}(\textbf{x}, t)$ and $\textbf{H}(\textbf{x}, t)$:
\begin{align}\label{eq:em_hamiltonian}
    \mathcal{H} =\frac{1}{2} \int_\Omega \bigg[ \varepsilon_0 \varepsilon(\textbf{x}) \textbf{E}^2(\textbf{x}, t) + \mu_0 \textbf{H}^2(\textbf{x}, t)\bigg] \textrm{d}^3\textbf{x}
\end{align}
where $\Omega$ denotes the spatial support of the electromagnetic fields and the square of a vector is assumed to be sum of squares of its components $\textbf{v}^2 = v_x^2 + v_y^2 + v_z^2$. For a given electromagnetic structure, this hamiltonian and the electromagnetic fields can be expressed in terms of a set of canonically conjugate variables  $\{ (p_i, q_i), i \in \mathbb{N}\}$ or functions $\{p_i(\zeta), q_i(\zeta), i \in \mathbb{N} \}$, where $\zeta$ is a continuous index, such that the dynamics predicted by Maxwell's equations is reproduced by the hamilton's equations of motion:
\begin{align}\label{eq:hamilton_eq}
    \partial_t p_i = -\frac{\delta \mathcal{H}}{\delta q_i} \text{ and } \partial_t q_i = \frac{\delta \mathcal{H}}{\delta p_i}
\end{align}
where the derivative is to be interpreted as a functional derivative if $p, q$ are functions. According to the second quantization principle \cite{peskin1995introduction}, the quantum description of the electromagnetic field can then be obtained by promotion $p$ and $q$ to operator $\hat{p}$ and $\hat{q}$, with the commutators given by the following rules:
\begin{enumerate}
    \item If ($p_i, q_i$) are scalar variables, then:
    \begin{align}
        [\hat{q}_i, \hat{q}_j] = 0, [\hat{p}_i, \hat{p}_j] = 0, [\hat{q}_i, \hat{p}_j] = \textrm{i}\hbar\delta_{i,j}
    \end{align}
    \item If ($\hat{p}_i(\zeta), \hat{q}_i(\zeta)$) are functions of the index $\zeta$ then:
    \begin{align}
        [\hat{q}_i(\zeta), \hat{q}_j(\zeta')] = 0, [\hat{p}_i(\zeta), \hat{p}_j(\zeta')] = 0, [\hat{q}_i(\zeta), \hat{p}_j(\zeta')] =\textrm{i} \hbar\delta_{i,j}\delta(\zeta - \zeta')
    \end{align}
\end{enumerate}
Expression for electric and magnetic field operators are assumed to be the same as their classical counterparts, with $(p_i, q_i)$ replaced with the corresponding operators. In the following subsections, we illustrate the application of this simple principle for three different electromagnetic structures --- lossless electromagnetic cavity, lossless waveguide and bulk medium.

\subsection{Lossless cavity}
For the purposes of this section, we assume that the lossless cavity is formed by enclosing a permittivity distribution $\varepsilon(\textbf{x})$ with a perfect electric conductor. The spatial support of the electromagnetic fields $\Omega$ will simply be the interior of the cavity. It is well known that the electromagnetic fields inside the cavity can be described entirely in terms of its modes $\textbf{E}_n(\textbf{x})$ and their resonant frequencies $\omega_n$ which satisfy the eigen-value equations:
\begin{align}
    \nabla \times \nabla \times \textbf{E}_n(\textbf{x}) = \mu_0\varepsilon_0 \omega_n^2 \varepsilon(\textbf{x})\textbf{E}_n(\textbf{x}) \ \forall \ \textbf{x} \in \Omega
\end{align}
along with the boundary conditions that:
\begin{align}
    \textbf{n}(\textbf{x}) \times \textbf{E}_n(\textbf{x}) = 0 \ \text{and} \ \textbf{n}(\textbf{x}) \times \textbf{H}_n(\textbf{x})  = 0 \ \forall \ \textbf{x} \in \partial \Omega
\end{align}
where $\textbf{H}_n(\textbf{x}) = \nabla \times \textbf{E}_n(\textbf{x}) / \textrm{i}\mu_0\omega_n$ is the magnetic field of the mode and $\partial\Omega$ is the boundary of $\Omega$. Note that since $\nabla \times \nabla \times$ is a perfectly real and positive-semidefinite operator, we can choose $\textbf{E}_n(\textbf{x})$ to be completely real. Moreover, they also satisfy the orthonormality conditions:
\begin{align}\label{eq:efield_ortho}
    \varepsilon_0\int_\Omega \varepsilon(\textbf{x})\textbf{E}_n(\textbf{x})\cdot \textbf{E}_m(\textbf{x}) \textrm{d}^3\textbf{x} = \delta_{n, m}
\end{align}
This orthonormality condition can also be equivalently be expressed in terms of the magnetic field $\textbf{H}_n(\textbf{x})$:
\begin{align}\label{eq:hfield_ortho}
    \mu_0\int_\Omega \textbf{H}_n(\textbf{x}) \cdot \textbf{H}_m(\textbf{x}) \textrm{d}^3\textbf{x} = -\delta_{n, m}
\end{align}
A general electromagnetic field inside the cavity can be expressed as a linear combination of the modal fields:
\begin{align}\label{eq:cavity_expansion}
    \textbf{E}(\textbf{x}, t) = \sum_{n} \omega_n q_n(t) \textbf{E}_n(\textbf{x}) \ \text{and} \ \textbf{H}(\textbf{x}, t) = \textrm{i}\sum_{n} p_n(t) \textbf{H}_n(\textbf{x})
\end{align}
Note that the factor of $\textrm{i}$ in the magnetic field expansion is required to ensure that the magnetic field is completely real. The dynamics of the electromagnetic fields is completely captured by the dynamics of the variables $p_n(t)$ and $q_n(t)0$. From the Maxwell's curl equations in time-domain, we can derive a set of ordinary differential equations for $p_n(t)$ and $q_n(t)$:
\begin{align}
    &\nabla \times \textbf{E}(\textbf{x}) = -\mu_0 {\partial}_t \textbf{H}(\textbf{x}, t) \implies \partial_t p_n(t) = -\omega_n^2 q_n(t) \\
    &\nabla \times \textbf{H}(\textbf{x}) = \varepsilon_0 \varepsilon(\textbf{x}) \partial_t \textbf{E}(\textbf{x}, t) \implies \partial_t q_n(t) = p_n(t)
\end{align}
To see that $p_i$ and $q_i$ are canonically conjugate variables, we use Eqs.~\ref{eq:em_hamiltonian} and \ref{eq:cavity_expansion} to express $\mathcal{H}$ in terms of $(p_i, q_i)$ --- with a straightforward application of the orthonormality conditions (Eqs.~\ref{eq:efield_ortho} and \ref{eq:hfield_ortho}), the hamiltonian evaluates to:
\begin{align}
    \mathcal{H} = \frac{1}{2} \sum_n (p_n^2(t) + \omega_n^2 q_n^2(t))
\end{align}
It can easily be verified that with this hamiltonian, hamilton's equations of motion (Eq.~\ref{eq:hamilton_eq}) reproduce the dynamical equations implied by Maxwell's equations, proving that $p_i$ and $q_i$ are indeed canonically conjugate. With an application of the second quantization principle, we thus obtain the following hamiltonian for the cavity:
\begin{align}
    \hat{H} = \frac{1}{2} \sum_i (\hat{p}_i^2 + \omega_i^2 \hat{q}_i^2)
\end{align}
with the commutation relations:
\begin{align}
    [\hat{q}_i, \hat{q}_j] = 0, [\hat{p}_i, \hat{p}_j] = 0, [\hat{q}_i, \hat{p}_j] = \textrm{i} \delta_{i,j}
\end{align}
This hamiltonian can be recast in terms of the annihilation operators $\hat{a}_n = (\omega_n \hat{q}_n + \textrm{i} \hat{p}_n) / \sqrt{2\hbar \omega_n}$ to obtain:
\begin{align}
    \hat{H} = \sum_i \hbar \omega_i \hat{a}_i^\dagger \hat{a}_i
\end{align}
with commutators $[\hat{a}_i, \hat{a}_j^\dagger] = \delta_{i, j}$ and $[\hat{a}_i, \hat{a}_j] = 0$. Using the expansion Eq.~\ref{eq:cavity_exp} and the definition of $\hat{a}_i$, the electric field operator can be expressed as:
\begin{align}
    \hat{\textbf{E}}(\textbf{x}) = \sum_i (2\hbar \omega_i)^{1/2} (\hat{a}_i + \hat{a}_i^\dagger) \textbf{E}_i(\textbf{x}) = \sum_i \bigg(\frac{2\hbar \omega_i}{\varepsilon_0 \int_\Omega \varepsilon(\textbf{x})\textbf{E}_i^2(\textbf{x})\textrm{d}^3\textbf{x} } \bigg)^{1/2} (\hat{a}_i + \hat{a}_i^\dagger) \textbf{E}_i(\textbf{x})
\end{align}
wherein the second expression for $\hat{\textbf{E}}(\textbf{x})$ holds even if $\textbf{E}_i(\textbf{x})$ is not normalized according to Eq.~\ref{eq:efield_ortho}. For a single-mode cavity, the hamiltonian and electric field operator can be further simplified to:
\begin{align}
    \hat{H} = \hbar \omega_C \hat{a}^\dagger \hat{a} \ \text{and} \ \hat{\textbf{E}}(\textbf{x}) = \bigg(\frac{2\hbar \omega_C}{\varepsilon_0 \int_\Omega \varepsilon(\textbf{x})\textbf{E}_C^2(\textbf{x})\textrm{d}^3\textbf{x} } \bigg)^{1/2} (\hat{a} + \hat{a}^\dagger) \textbf{E}_C(\textbf{x})
\end{align}
\subsection{Lossless waveguide}
A procedure similar to that followed for quantizing a cavity can be followed for quantizing a waveguide. In this section, we restrict ourselves to waveguides that can be described completely by guided modes and have no radiation modes --- such a waveguide can be formed, for e.g., if a perfect electrical conducting pipe encloses a permittivity distribution $\varepsilon(\boldsymbol{\rho}) \equiv \varepsilon(y, z)$ which is independent of the propagation direction (assumed to be $x$). A waveguide can be completely characterized by propagating modes $\textbf{E}_i(\textbf{x}; \beta)$ which are of the form:
\begin{align}
    \textbf{E}_i(\textbf{x}; \beta) = \textbf{E}_i(\boldsymbol{\rho}; \beta) \exp(\textrm{i}\beta x)
\end{align}
where $\beta$ is the propagation constant. The frequency at which this mode propagates is given by a dispersion relation $\omega_n(\beta)$. It follows from Maxwell's equations that the modes satsify the orthonormality relations:
\begin{align}\label{eq:wg_efield_ortho}
    \varepsilon_0\int_\Omega \varepsilon(\boldsymbol{\rho})\textbf{E}_i^*(\textbf{x}; \beta) \cdot \textbf{E}_j(\textbf{x}; \beta') \textrm{d}^3\textbf{x} = \delta(\beta - \beta') \delta_{i, j}
\end{align}
where $\Omega = \{(x, \boldsymbol{\rho}) | \boldsymbol{\rho} \in \Gamma \}$ is the spatial support of the electromagnetic field --- this is a set of all points that lie within the waveguide cross section $\Gamma$ at any plane normal to the propagation direction. This orthonormality condition can be equivalently expressed in terms of the magnetic field of the mode ($\textbf{H}_i(\textbf{x}; \beta) = \nabla \times \textbf{E}_i(\textbf{x}; \beta) / \textrm{i}\mu_0\omega_i(\beta)$):
\begin{align}\label{eq:wg_hfield_ortho}
    \mu_0 \int_\Omega \textbf{H}_i^*(\textbf{x}; \beta) \cdot \textbf{H}_j(\textbf{x}; \beta') \textrm{d}^3\textbf{x} = \delta(\beta - \beta') \delta_{i, j}
\end{align}
An electromagnetic field inside the waveguide can then be expressed as a superposition of these propagating modes (we only consider modes propagating in $+x$ direction):
\begin{subequations}\label{eq:wg_exp}
\begin{align}
    \textbf{E}(\textbf{x}, t) = \sum_i \int_0^\infty \sqrt{2}\text{Re}[\{\textrm{i}p_i(t; \beta) + \omega_i(\beta) q_i(t; \beta)\} \textbf{E}_i(\textbf{x}; \beta)] \textrm{d}\beta \\
    \textbf{H}(\textbf{x}, t) = \sum_i \int_0^\infty \sqrt{2}\text{Re}[\{\textrm{i}p_i(t; \beta) + \omega_i(\beta) q_i(t; \beta)\} \textbf{H}_i(\textbf{x}; \beta)] \textrm{d}\beta
\end{align}
\end{subequations}
where $p_i(t;\beta)$ and $q_i(t; \beta)$ are real functions of time. Note that $\textrm{i} p_i(t; \beta) + \omega_i q_i(t; \beta)$ is the complex amplitude of the contribution of the mode indexed by $(i, \beta)$ in the full electromagnetic field, and thus oscillate at frequency $\omega_i(\beta)$:
\begin{align}
    \partial_t [\textrm{i} p_i(t; \beta) + \omega_i q_i(t; \beta)] = -\textrm{i}\omega_i[\textrm{i} p_i(t; \beta) + \omega_i q_i(t; \beta)] \implies \partial_t q_i(t; \beta) = p_i(t; \beta) \ \text{and } \partial_t p_i(t; \beta) = -\omega_i^2(\beta) q_i(t; \beta)
\end{align}
These constitute the dynamical equations for the waveguide fields. To see that this implies that $p_i(t; \beta)$ and $q_i(t; \beta)$ are canonically conjugate, we express the hamiltonian $\mathcal{H}$ in terms of $p_i(t; \beta)$ and $q_i(t;\beta)$ --- using Eqs.~\ref{eq:wg_exp}, and with an application of the orthonormality conditions (Eqs.~\ref{eq:wg_efield_ortho} and \ref{eq:wg_hfield_ortho}), we obtain:
\begin{align}
    \mathcal{H} = \frac{1}{2}\sum_{i} \int_0^\infty \big[ p_i^2(t;\beta) + \omega_i^2(\beta) q_i^2(t;\beta) \big]\textrm{d}\beta
\end{align}
It can easily verified that with this hamiltonian, hamilton's equations of motion (Eq.~\ref{eq:hamilton_eq}) reproduces the dynamical equations stated above, proving that $p_i(t; \beta)$ and $q_i(t; \beta)$ are canonically conjugate. Applying the second quantization principle, the quantum hamiltonian $\hat{H}$ can then be expressed:
\begin{align}
    \hat{H} = \sum_{i} \int_0^\infty [\hat{p}_i^2(\beta) + \omega_i^2 \hat{q}_i^2(\beta)] \textrm{d}\beta
\end{align}
with the commutation relations
\begin{align}
    [\hat{q}_i(\beta), \hat{q}_j(\beta')] = 0, [\hat{p}_i(\beta), \hat{p}_j(\beta')] = 0, [\hat{q}_i(\beta), \hat{p}_j(\beta')] = \textrm{i}\delta(\beta - \beta')\delta_{i,j}
\end{align}
Similar to that done for optical cavities, we can define an annihilation operator $a_i(\beta)$ for each waveguide mode via $\hat{a}_i(\beta) = [\omega_i(\beta) \hat{q}_i(\beta) + \textrm{i}\hat{p}_i(\beta)] / \sqrt{2\hbar \omega_i(\beta)}$, in terms of which the hamiltonian reduces to:
\begin{align}
    \hat{H} = \sum_i \int_0^\infty \hbar\omega_i(\beta) \hat{a}_i^\dagger(\beta) \hat{a}_i(\beta) \textrm{d}\beta
\end{align}
with the commutation relations $[\hat{a}_i(\beta), \hat{a}_j(\beta')] = 0$ and $[\hat{a}_i(\beta), \hat{a}_j^\dagger(\beta')] = \delta_{i, j}\delta(\beta - \beta')$. The electric field operator can be expressed as:
\begin{align}
    \hat{\textbf{E}}(\textbf{x}) &= \sum_{i}\int_{0}^\infty [2\hbar \omega_i(\beta)]^{1/2} \hat{a}_i(\beta) \textbf{E}_i(\boldsymbol{\rho}; \beta)\exp(-\textrm{i}\beta z) \textrm{d}\beta + \text{h.c.} \\
    &= \sum_{i}\int_{0}^\infty \bigg[\frac{2\hbar \omega_i(\beta)}{\int_\Gamma \varepsilon(\boldsymbol{\rho})|\textbf{E}_i(\boldsymbol{\rho}; \beta)|^2 \textrm{d}^2 \boldsymbol{\rho}} \bigg]^{1/2} \hat{a}_i(\beta) \textbf{E}_i(\boldsymbol{\rho}; \beta)\exp(-\textrm{i}\beta z) \textrm{d}\beta + \text{h.c.}
\end{align}

\section{Analysis of interferometers}\label{sec:interferometers}
\subsection{Linear optical elements in loss channels}
In this subsection, we describe the hamiltonians and derive the heisenberg equations of motion for linear optical elements that act on light propagating in an ideal loss channel. The two linear optical elements that we consider here are the phase shifter and the beam splitter --- almost all, more complicated, linear optical elements can be decomposed into their cascades.

A \emph{phase shifter} is a linear optical element that imparts a constant phase to the light propagating through it. Physical realizations of phase shifters are often as simple as just adding extra optical path lengths to the incident field, or using thermal or electro-optical effects to change the local refractive index of the loss channel. Here we focus on broadband phase shifters --- phase shifters that impart the same frequency to all the frequencies propagating in the loss channel. The hamiltonian for broadband phase shifter on a loss channel with frequency annihilation operator $a(\omega)$ and spatial annihilation operator $a(x)$ is given by:
\begin{align}
    \hat{H} = \int_{-\infty}^\infty \omega \hat{a}(\omega)^\dagger \hat{a}(\omega) \textrm{d}\omega + v_G V \hat{a}^\dagger(x=0)\hat{a}(x=0)
\end{align}
where $V$ is the `strength' of the phase shifter (this is related to the phase shift below) and it is assumed that the phase shifter acts at $x = 0$ on the loss channel. Note that:
\begin{align}
    \int \omega \hat{a}^\dagger(\omega) \hat{a}(\omega) \textrm{d}\omega = \int_{-\infty}^\infty \omega \hat{a}^\dagger(x)\hat{a}(\omega) \exp\bigg(-\textrm{i}\frac{\omega x}{v_G}\bigg)\frac{\textrm{d} x\textrm{d}\omega}{\sqrt{2\pi v_G}} = -\textrm{i}v_G \int_{-\infty}^\infty \hat{a}^\dagger(x) \partial_x \hat{a}(x) \textrm{d}x
\end{align}
where
\begin{align}
    \partial_x \hat{a}(x) = \frac{\textrm{i}}{v_G}\int_{-\infty}^\infty  \hat{a}(\omega)\exp\bigg(\frac{\textrm{i}\omega x}{v_G} \bigg) \frac{\textrm{d}\omega}{\sqrt{2\pi v_G}}
\end{align}
with which the phase shifter hamiltonian can be fully expressed in terms of the position annihilation operator $a(x)$:
\begin{align}
    H = -\textrm{i}v_G \int_{-\infty}^\infty \hat{a}^\dagger(x) \partial_x \hat{a}(x)\textrm{d}x + v_G V \hat{a}^\dagger(x = 0)\hat{a}(x=0)
\end{align}
To analyze this system, we employ the heisenberg picture --- using the commutators for the position annihilation operators allows a straightforward calculation of the equation of motion for the heisenberg picture annihilation operator $a(t; x)$:
\begin{align}\label{eq:heisenberg_pshifter}
    \frac{1}{v_G} \partial_t a(t; x) + \partial_x a(t; x) = -\textrm{i}V a(t; x=0)\delta(x)
\end{align}
which we need to solve under the initial condition that $a(0; x)$ is identical to the Schrondinger picture operator $a(x)$. Note that in the absence of the phase shifter (i.e. $V = 0$), this equation can be trivially solved to obtain $a(t; x) = a(x-v_Gt)$ which correspond to the wave packet in the loss channel propagating along the $x$ direction with velocity $v_G$. For $V \neq 0$, a general solution to this equation can be written as:
\begin{align}\label{eq:pshifter_form}
    a(t; x) =
    \begin{cases}
    a(x - v_G t) & \text{ if } x < -v_G t \\
    S a(x - v_G t) & \text{ if } -v_Gt \leq x \leq 0 \\
    a(x - v_G t) & \text{ if } x > 0
    \end{cases}
\end{align}
where $S$ is a scalar that captures the impact of the phase shifter on the loss channel. This form of the solution for the dynamical equations can easily be physically interpreted by noting that excitations in the loss channel propagate along the $+x$ direction at a speed $v_G$. Since excitations that reach a point in the region $x < -v_G t$ and $x > 0$ at time $t$ wouldn't have encountered the beam splitter (which as at $x = 0$), they are simply described by translating the initial operator $a(x)$ in time. Excitations that reach a point in the region $-v_G t \leq x \leq 0$ at time $t$ would have encountered the beam splitter at $t = 0$ which performed an (unknown) linear operation described by $S$ on the operators. To calculate $S$, we integrate Eq.~\ref{eq:heisenberg_pshifter} across an infinitesmal interval around $x = 0$ to obtain:
\begin{align}
    a(t; x = 0^+) - a(t; x = 0^-) = -\frac{\textrm{i}V}{2}[a(t; x = 0^+) + a(t; x = 0^-)]
\end{align}
Using this along with Eq.~\ref{eq:pshifter_form}, we immediately obtain:
\begin{align}
    S = \frac{1 - \textrm{i} V / 2}{1 + \textrm{i} V/2} = \exp(\textrm{i} \varphi)
\end{align}
where $\varphi = 2 \text{tan}^{-1}(V/2)$ is the phase-shift induced by the phase-shifter.
 
A \emph{beam-splitter} is a very commonly used linear optical device that interferes two propagating optical signals --- physical realization of a beam splitter typically used are a partially transmitting mirror for free-space optical beams, and a directional coupler for optical waveguides. While the exact dynamics of a beam splitter can be very complicated, if we assume that the bandwidth of the beam splitter is much larger than the bandwidth of the optical signals that it is interfering, it can be analyzed with a simple model, which we describe in this section. Consider two loss channels, with frequency annihilation operators $a_\omega, b_\omega$ and spatial annihilation operators $a_x, b_x$ --- the hamiltonian for a broadband beam splitter in between these two loss channels is given by:
\begin{align}
H = \int_{-\infty}^{\infty} \omega a_\omega^\dagger a_\omega \textrm{d}\omega + \int_{-\infty}^\infty \omega b_\omega^\dagger b_\omega \textrm{d}\omega + \textrm{i}v_g (V^* b_{x=0}^\dagger a_{x=0} - V a_{x=0}^\dagger b_{x=0})
\end{align}
where we assume that the beam splitter couples the two loss channels at $x = 0$, and $V$ is a dimensionless constant that governs the strength of this coupling. As was done for the phase shifter, this hamiltonian can be expressed completely in terms of the position annihilation operator:
\begin{align}
    H = -\textrm{i}v_G \bigg[\int_{-\infty}^{\infty} a_x^\dagger \partial_x a_x \textrm{d}x + \int_{-\infty}^\infty b_x^\dagger \partial_x b_x \textrm{d}x \bigg] + \textrm{i}v_G (V^* b_{x=0}^\dagger a_{x=0} - V a_{x=0}^\dagger b_{x=0})
\end{align}
To analyze this system, we employ the heisenberg picture --- using the commutators for the position annihilation operators allows a straightforward calculation of the equations of motion for the heisenberg picture annihilation operators $a(t; x)$ and $b(t; x)$:
\begin{subequations}\label{eq:beam_splitter_dyn}
\begin{align}
    &\frac{1}{v_G} \partial_t a(t; x) + \partial_x a(t; x) =  -V \delta(x) b(t; 0) \\
    &\frac{1}{v_G} \partial_t b(t; x) + \partial_x b(t; x) =  V^* \delta(x) a(t; 0)
\end{align}
\end{subequations}
which we need to solve under the initial condition that $a(0; x)$ and $b(0; x)$ are identical to the Schrodinger picture operator $a(x)$ and $b(x)$. Note that in the absence of the beam splitter (i.e.~$V = 0$), these equations can be trivially solved to obtain $a(t; x) = a(x - v_G t)$ and $b(t; x) = b(x - v_G t)$ which correspond to the propagation of a wave packet down the two waveguides at velocity $v_G$ without coupling to each other. For $V \neq 0$, a general solution to these equations can be written as:
\begin{align}
    a(t; x) = \begin{cases} 
        a(x - v_G t) \ & \text{ if } x < - v_G t \\
        S_{a, a} a(x - v_Gt) + S_{a, b} b(x - v_G t) \ & \text{ if } -v_G t \leq x \leq 0 \\
        a(x - v_G t) \ & \text{ if } x > 0
    \end{cases}
\end{align}
\begin{align}
    b(t; x) = \begin{cases} 
        b(x - v_G t) \ & \text{ if } x < - v_G t \\
        S_{b, a} a(x - v_Gt) + S_{b, b} b(x - v_G t) \ & \text{ if } -v_G t \leq x \leq 0 \\
        b(x - v_G t) \ & \text{ if } x > 0
    \end{cases}
\end{align}
where $S_{i, j} \ \forall \ i, j \in \{a, b\}$ are coefficients of the scattering matrix for the beam splitter, which we still have to compute. To determine these coefficients, we integrate with respect to $x$ Eq.~\ref{eq:beam_splitter_dyn} across an infinitesmal region around $x=0$ to obtain:
\begin{subequations}
\begin{align}
&a(t; x = 0^+) - a(t; x = 0^-) = \frac{V}{2} [b(t; x = 0^+) + b(t; x = 0^-)] \\
&b(t; x = 0^+) - b(t; x = 0^-) = -\frac{V^*}{2}[a(t; x = 0^+) + a(t; x = 0^-)]
\end{align}
\end{subequations}
Using the expressions for $a(t; x)$ and $b(t; x)$, these equations can be translated to a set of linear equations in the scalars $S_{a, a}, S_{a, b}, S_{b, a}$ and $S_{b, b}$ which can be solved to obtain the following expression for the beam-splitter scattering matrix:
\begin{align}
    S = \begin{bmatrix} S_{a,a} & S_{a, b} \\
                        S_{b,a} & S_{b,b} \end{bmatrix} =
        \frac{1}{1 + |V|^2/4}\begin{bmatrix} 1 - |V|^2/4 & V \\
            -V^* & 1 - |V|^2/4
        \end{bmatrix}
\end{align}
The conventional definition of a beam splitter assumes that $V$ is purely real, and defining $\theta$ via $\sin \theta = V / (1 + |V|^2/4)$ and $\cos \theta = (1 - |V|^2/4) / (1 + |V|^2/4)$, we obtain:
\begin{align}
        S = \begin{bmatrix} \cos \theta & -\sin \theta \\
            \sin \theta & \cos \theta
        \end{bmatrix}
\end{align}
In particular, a 50-50 beam splitter is defined as a beam splitter with $\theta = \pi / 4$ --- this interferes both the input ports equally, constructively in one output arm and destructively in the other.
\subsection{Analysis of Hanbury-brown Twiss and Hong-Ou Mandel interferometers}
\noindent Here, we analyze the two interferometers shown in Fig.~\ref{figure:interferometer}. Note that both the interferometers apply the same optical transformation to the input fields --- in one case, one of the arms have vacuum state as an input, whereas in the other case both arms are excited with the light source being characterized. For our analysis, we assume that the photon pulses assumed by the light have a pulse width that is much smaller than the other optical lengths involved in the setup (e.g. the distance between the sources or detectors from the phase shifters or beam splitters). In this case,
\begin{subequations}\label{eq:input_output_interferometer}
\begin{align}
    a(t; L) = \frac{a(L - v_G t) - \exp(\textrm{i}\varphi) b(L - v_G t)}{\sqrt{2}} \\
    b(t; L) = \frac{a(L - v_G t) + \exp(\textrm{i}\varphi) b(L - v_G t)}{\sqrt{2}}
\end{align}
\end{subequations}
where we assume that both $t$ are large enough for the light emitted by the sources to have propagated to the detectors --- if this is not the case, then $a(t; L)$ or $b(t; L)$ will simply annihilate the state $\ket{\Psi}$ emitted by the sources since $L - v_G t$ would lie outside the spatial region corresponding to the photon pulse. Consider now the expectation $\mathbb{E}_\varphi[\bra{\Psi} a(t_1; L) b(t_2; L) b^\dagger(t_2; L) a^\dagger(t_1; L) \ket{\Psi}]$ --- using Eq.~\ref{eq:input_output_interferometer} and explicitly averaging over $\varphi$:
\begin{align}\label{eq:decomposition_two_time}
    \mathbb{E}_\varphi&\big[\bra{\Psi} a(t_1; L) b(t_2; L) b^\dagger(t_2; L) a^\dagger(t_1; L)  \ket{\Psi}\big] =\nonumber\\ \frac{1}{4}\bigg[&\sum_{v\in\{a, b\}}\bra{\Psi_v}v^\dagger(L-v_G t_1) v^\dagger(L-v_G t_2) v(L-v_G t_1) v(L - v_G t_2)\ket{\Psi_v} + \nonumber \\
    & \sum_{(u, v)\in \mathcal{S}} \bra{\Psi_v} v^\dagger(L - v_G t_1) v(L - v_G t_1)  \ket{\Psi_v}\bra{\Psi_u} u^\dagger(L - v_G t_2) u(L - v_G t_2)  \ket{\Psi_u} - \nonumber \\
    & \sum_{(u, v) \in \mathcal{S}} \bra{\Psi_v}v^\dagger(L-v_G t_1) v(L - v_G t_2) \ket{\Psi_v} \bra{\Psi_u}u^\dagger(L-v_G t_2) u(L - v_G t_1) \ket{\Psi_u} \bigg]
\end{align}
where $\mathcal{S} = \{(a, b), (b, a)\}$ and we have assumed that the two sources are not entangled with each other --- note that if both the sources emit a pure state, then this would imply that the state of the composite system is $\ket{\Psi_a}\ket{\Psi_b}$ and if the sources emit a mixed state, then the density matrix of the composite system is $\hat{\rho}_a \hat{\rho}_b$. Consequently, the expectation of a product of $\hat{a}(x)$ and $\hat{b}(x)$ operators (and their conjugates) can be factorized:
\begin{align}
    \bra{\Psi} f_a[a(x), a^\dagger(x)] f_b[b(x), b^\dagger(x)] \ket{\Psi} = \bra{\Psi_a} f_a[a(x), a^\dagger(x)] \ket{\Psi_a} \bra{\Psi_b}f_b[b(x), b^\dagger(x)]\ket{\Psi_b}
\end{align}
Similar expressions for the expections $\mathbb{E}_\varphi(\bra{\Psi} a(t; L) a^\dagger(t; L) \ket{\Psi})$ and $\mathcal{E}_\varphi(\bra{\Psi} b(t; L) b^\dagger(t; L) \ket{\Psi}$ can be obtained by using Eq.~\ref{eq:input_output_interferometer}:
\begin{align}\label{eq:decomposition_one_time}
\mathbb{E}_\varphi(\bra{\Psi}a^\dagger(t; L) a(t; L) \ket{\Psi}) = \mathbb{E}_\varphi[\bra{\Psi} b^\dagger(t; L) b(t; L) \ket{\Psi}] 
= \frac{1}{2}\bigg[\sum_{v \in \{a, b\}}\bra{\Psi_v} v^\dagger(L - v_G t) v(L - v_Gt)\ket{\Psi_v}\bigg]
\end{align}

With the help of these relationships, we can now prove the results about the two interferometers stated in section . Consider the Hanbury-Brown Twiss interferometer --- since the second waveguide (labelled by $b(x)$) is in vacuum state, it follows that:
\begin{subequations}
\begin{align}
    &\mathbb{E}_\varphi \big[\bra{\Psi} a^\dagger(t_1; L) b^\dagger(t_2; L) b^\dagger(t_2; L) a(t_1; L)  \ket{\Psi}\big] = \frac{1}{4}\bra{\Psi_a}a^\dagger(L-v_G t_1) a^\dagger(L-v_G t_2) a(L-v_G t_1) a(L - v_G t_2)\ket{\Psi_a} \\
    &\mathbb{E}_\varphi[\bra{\Psi}a^\dagger(t; L) a(t; L) \ket{\Psi}] = \mathbb{E}_\varphi[\bra{\Psi} b^\dagger(t; L) b(t; L) \ket{\Psi}] = \frac{1}{2}\bra{\Psi_a} a^\dagger(L - v_G t) a(L - v_Gt)\ket{\Psi_a}
\end{align}
\end{subequations}
from which it immediately follows that $g^{(2)}_\text{HBT}(t_1, t_2)$ and $g^{(2)}_\text{HBT}[0]$ are exactly equal to $g^{(2)}(t_1, t_2)$ and $g^{(2)}[0]$.

Similarly, consider the Hong-Ou Mandel interferometer \cite{Fischer2016} --- assuming that both the sources emit a single photon state (pure or mixed), it immediately follows that the second order correlations in the sources vanish: $\bra{\Psi_v}v^\dagger(L-v_G t_1) v^\dagger(L-v_G t_2) v(L-v_G t_1) v(L - v_G t_2)\ket{\Psi_v} = 0$ for $v \in \{a, b\}$. Using the general form for the density matrix of the light emitted by a single photon source, the remaining expectations can be immediately evaluated:
\begin{align}
    \bra{\Psi_v} v^\dagger(L - v_G t_1) v(L - v_G t_2) \ket{\Psi_v} = P_{1, v} \rho^{(1)}_v(L - v_G t_2, L - v_G t_1) \ \text{for } v \in \{a, b\}
\end{align}
We can now evaluate the integrals appearing in the definition of $g^{(2)}_\text{HOM}[0]$ --- in particular note that:
\begin{subequations}
\begin{align}
    &\int_{-\infty}^\infty \bra{\Psi_v} v^\dagger(L - v_G t) v(L - v_G t) \ket{\Psi_v} \textrm{d}t = v_G P_{1, v} \\
    &\int_{-\infty}^\infty \int_{-\infty}^\infty \bra{\Psi_v} v^\dagger(L - v_G t_1) v(L - v_G t_2) \ket{\Psi_v}\bra{\Psi_u} u^\dagger(L - v_G t_2) u(L - v_G t_1) \ket{\Psi_u} \textrm{d}t_1 \textrm{d}t_2 \nonumber \\ & =v_G^2 P_{1, v} P_{1, u} \int \int \rho^{(1)}_v(x_1, x_2) \rho^{(1)}(x_2, x_1) \textrm{d}x_1 \textrm{d} x_2 \nonumber = v_G^2 P_{1, v} P_{1, u} \int \int \rho^{(1)}_v(x_1, x_2) \rho^{(1)*}_u(x_1, x_2) \textrm{d}x_1 \textrm{d}x_2\end{align}
\end{subequations}
With these expressions, one readily obtains:
\begin{align}
    g^{(2)}_\text{HOM}[0] = \frac{2 P_{1, a}P_{1, b}}{(P_{1, a} + P_{1, b})^2}\bigg[1 - \int_{-\infty}^{\infty} \int_{-\infty}^{\infty} \rho^{(1)}_a(x_1, x_2) \rho^{(1)*}_b(x_1, x_2) \textrm{d}x_1 \textrm{d}x_2 \bigg]
\end{align}
\end{document}